\documentclass[prb, twocolumn, floatfix, superscriptaddress, longbibliography, reprint]{revtex4-2}
\usepackage[pdftex,plainpages=false,colorlinks=true,linkcolor=blue, citecolor=blue, urlcolor=blue]{hyperref}
\usepackage{amsfonts}
\usepackage{amsmath}
\usepackage{amssymb}
\usepackage{graphicx}% Include figure files
\usepackage{natbib}
\usepackage{color}
\usepackage{placeins}
\usepackage{physics}

\begin{document}
 
\title{Charge gap and charge redistribution among copper and oxygen orbitals in the normal state of the Emery model}
\author{G. L. Reaney}
\affiliation{Department of Physics, Royal Holloway, University of London, Egham, Surrey, UK, TW20 0EX}
\author{N. Kowalski}
\affiliation{D\'epartement de physique, Institut quantique \& RQMP, Universit\'e de Sherbrooke, Sherbrooke, Qu\'ebec, Canada J1K 2R1}
\author{A.-M. S. Tremblay}
\affiliation{D\'epartement de physique, Institut quantique \& RQMP, Universit\'e de Sherbrooke, Sherbrooke, Qu\'ebec, Canada J1K 2R1}
\author{G. Sordi}
\email[corresponding author: ]{giovanni.sordi@rhul.ac.uk}
\affiliation{Department of Physics, Royal Holloway, University of London, Egham, Surrey, UK, TW20 0EX}
\date{\today}

\begin{abstract}
Unraveling the behavior of the electrons in the copper-oxygen planes of cuprate superconductors remains a challenge. Here we examine the electronic charge redistribution among planar copper and oxygen orbitals and the charge gap using the Emery model in the normal state, solved with cellular dynamical mean-field theory at finite temperature. We quantify the charge redistribution as a function of the onsite Coulomb repulsion on the copper orbitals, the bare copper-oxygen energy difference, and the hole or electron doping. We find that the position relative to the metal to insulator boundary of the Zaanen-Sawatzky-Allen diagram determines the charge redistribution among copper and oxygen orbitals. For a fixed bare Cu-O energy difference, an increase in the Cu electron repulsion leads to a transfer of the electronic charge from Cu to O orbitals. For a fixed charge gap size of the undoped state, as the system evolves from a charge-transfer to a Mott-Hubbard regime, the electronic charge is transferred from Cu to O orbitals. 
Our findings posit the Coulomb repulsion and the bare charge-transfer energy as key drivers of the microscopic process of charge redistribution in the CuO$_2$ plane. They quantify the anticorrelation between the charge gap size and oxygen hole content. They show that for fixed band-structure parameters, the charge gap and the charge redistribution between Cu and O orbitals provide a way to understand observed trends in cuprates. 
\end{abstract}
 
\maketitle

\section{Introduction}

Cuprate superconductors are layered materials with copper-oxygen planes separated by layers of ions. These layers can dope the planes with electron or holes, thereby redistributing the charge between planar copper and oxygen orbitals~\cite{keimerRev, Dagotto:RMP1994, ift}. Experimental work found an intriguing correlation between the charge redistribution in the copper-oxygen plane and the maximum superconducting critical temperature $T_c^{\rm max}$, with $T_c^{\rm max}$ increasing when the electronic charge is transferred from the oxygen to the copper orbitals~\cite{Rybicki:NatComm2016, Jurkutat:PNAS2023}. This finding underscores the key role of the charge redistribution among planar copper and oxygen orbitals in driving the superconducting pairing~\cite{Rybicki:NatComm2016, Nicolas:PNAS2021, Davis:PNAS2022, Jurkutat:PNAS2023}. 
However, the strong electron-electron correlations on the copper orbitals~\cite{Dagotto:RMP1994} make it challenging to decipher the redistribution of the electrons in the copper-oxygen plane. 

Here we address this challenge by studying the dependence of the charge redistribution in the copper-oxygen plane on the strength of the electron-electron repulsion on the copper orbitals and on the bare copper-oxygen energy difference. We do that using the Emery model~\cite{Emery_1987, Varma_1987}, which, in its simplest form, comprises three orbitals per unit cell -- two oxygen $2p_x, 2p_y$ orbitals hybridised with one copper $3d_{x^2-y^2}$ orbital with an onsite electron-electron repulsion (see Fig.~\ref{fig:EmeryModel}(a)). 
For an electronic density of five electrons, or equivalently one hole, per CuO$_2$ unit cell, the Emery model can show different correlated insulating states. 
According to the Zaanen-Sawatzky-Allen scheme~\cite{zsa}, the ratio of the Cu onsite repulsion $U_d$ and the charge-transfer energy $\Delta$ (the energy difference between the upper Hubbard band and the onsite energy of the O orbitals) sets the Emery model in the charge-transfer regime for $U_d \gtrsim \Delta$, in the Mott-Hubbard regime for $U_d \lesssim \Delta$, or in the intermediate regime between the two for $U_d \approx \Delta$ (see sketch of the density of states in Fig.~\ref{fig:EmeryModel}(b,c)). 
Strictly speaking, the Emery model was proposed in the charge-transfer regime~\cite{Emery_1987}. This is the regime that is appropriate for cuprate superconductors, as we will confirm quantitatively in this paper. To gain insights into general trends, we also explore the Emery model in the Mott-Hubbard regime.  

Due to the Cu-O hybridization, the hole per CuO$_2$ unit cell that becomes localised in these correlated insulators has a mixed $d$-$p$ character, or, equivalently, is shared between Cu and O orbitals. 
The mixed $d$-$p$ character of this hole is further controlled by the onsite repulsion on the Cu orbital and by the energy difference between the Cu and O orbitals. 
The first goal of this work is to understand and quantify the orbital character of this hole at zero doping, i.e. how this hole is shared among the copper and oxygen orbitals, when the system evolves from the charge-transfer to the Mott-Hubbard insulating regime. 
The second goal of this work is to understand and quantify the effect of electron and hole doping on the charge redistribution among Cu and O orbitals when the system evolves from the charge-transfer to the Mott-Hubbard regime. 
We analyse the charge redistribution at zero and finite electron/hole doping with the Cu and O occupancies. This method enables a comparison with the experimental results on cuprates obtained by nuclear magnetic resonance (NMR) in Refs.~\cite{Rybicki:NatComm2016, Jurkutat:PNAS2023}.

Owing to the challenge posed by the strong electron-electron correlations on the Cu orbitals in these regimes, to reach these goals we solve the Emery model with finite temperature cellular extension~\cite{maier, kotliarRMP, tremblayR} of dynamical mean-field theory~\cite{rmp}. This is a nonpertubative method allowing us to treat both temporal and spatial electronic fluctuations. Our key strategy is to construct the Zaanen-Sawatzky-Allen diagram $U_d$ vs $\Delta$ of the Emery model. The main finding of our work is that the position relative to the metal-insulator boundary of the Zaanen-Sawatzky-Allen diagram controls the redistribution of the charges among Cu and O orbitals. 
First, at fixed energy difference between Cu and O orbitals, an increasing $U_d$ leads to an increase of Cu hole content at the expenses of the O hole content, for either the Mott-Hubbard or the charge-transfer regime. This is because $U_d$ induces an effective suppression of the charge fluctuations between Cu and O orbitals. Second, for a fixed charge gap size, as the Emery model evolves from the charge-transfer to the Mott-Hubbard regime, the Cu hole content increases at the expenses of the O hole content. This is due to the holes that become more localised on the Cu orbitals. 

These findings have three main implications. First, they posit $U_d$ and $\Delta$ as possible key microscopic mechanisms governing the redistribution of charges among Cu and O orbitals in the CuO$_2$ plane. 
Second, they enable us to quantify the anticorrelation between the size of the correlated gap and the O hole content, all the way from the charge-transfer to the Mott-Hubbard regime, supporting earlier discussions~\cite{Nicolas:PNAS2021, Davis:PNAS2022, Jurkutat:PNAS2023}. 
Third, they provide a framework for connecting microscopic model parameters to some trends in cuprate phenomenology related to the charge redistribution in the CuO$_2$ plane and to the size of the charge-transfer gap. More specifically, we reveal that the intersection between O hole content isolines and charge gap isolines in the Zaanen-Sawatzky-Allen diagram sets the systems with larger O hole content deeper in the charge-transfer regime, i.e. in regions of higher $U_d$ and smaller $\Delta$. 

The rest of the paper is organised as follows. Section~\ref{sec:model} summarises our model and method. Section~\ref{sec:zerodoping} examines how electrons are shared among Cu and O orbitals at zero doping. Then, Section~\ref{sec:finitedoping} analyses the electronic charge redistribution in the CuO$_2$ plane upon electron or hole doping. Then, in Section~\ref{sec:discussion} we show that the Zaanen-Sawatzky-Allen diagram is a good framework to rationalise our results, and in Section~\ref{sec:conclusion} we summarise our main findings.

\section{Model and Method}
\label{sec:model}

\begin{figure}[ht!]
\centering{
\includegraphics[width=1.0\linewidth]{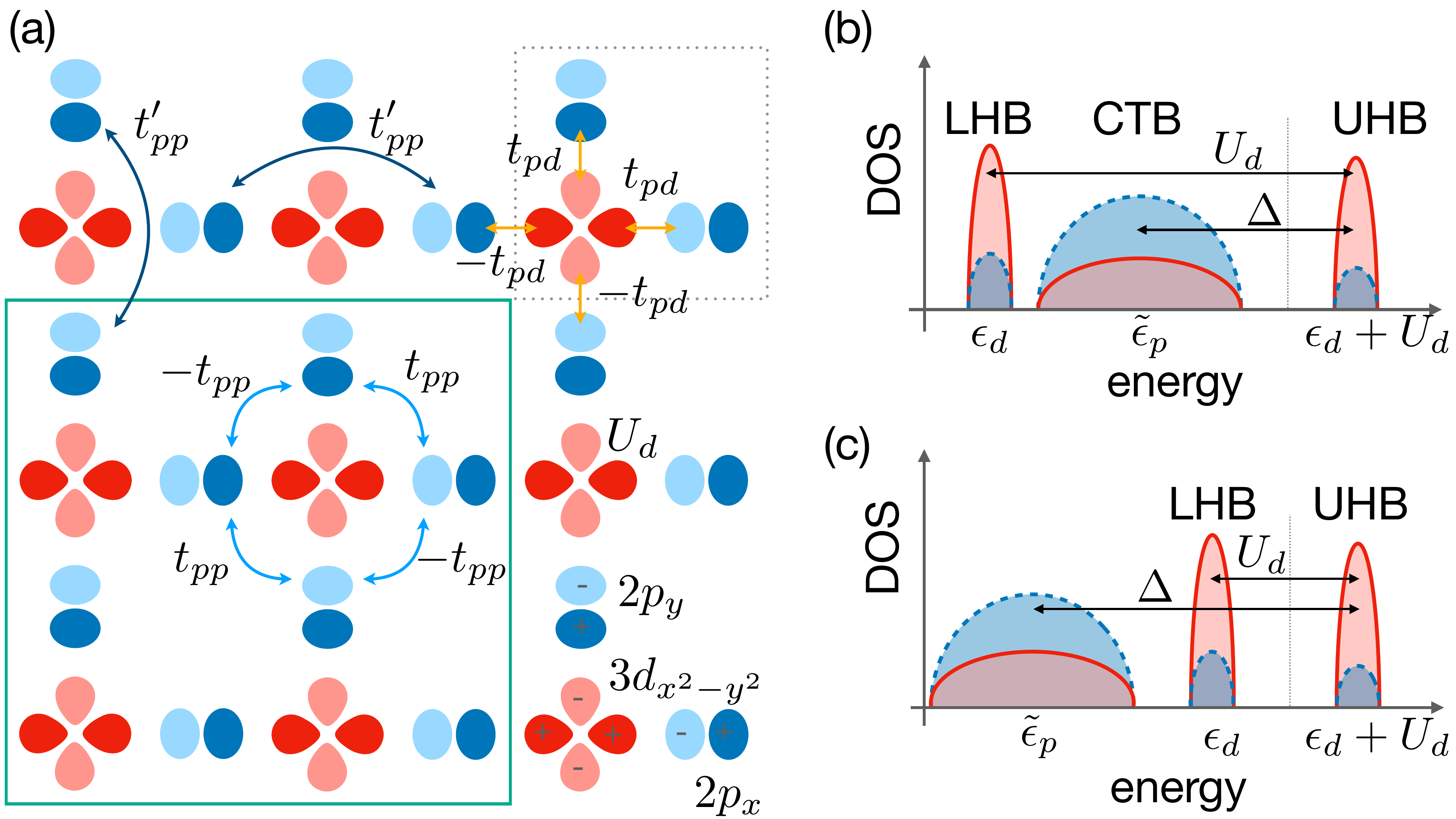}}
\caption{(a) Sketch of Emery model. The CuO$_2$ unit cell (dotted grey square), consists of a Cu $3d_{x^2-y^2}$ orbitals (red) of onsite energy $\epsilon_d$ and the O $2p_x$, $2p_y$ orbitals (blue) of onsite energy $\epsilon_p$. The positive (negative) sign of the orbital lobes are denoted by bright (pale) colors.
We use the convention that the overlap between orbitals of opposite (equal) sign leads to a positive (negative) hopping amplitude. $t_{pd}$ is the hopping amplitude between the nearest-neighbor Cu-O orbitals (orange arrows). $t_{pp}$ and $t_{pp}^\prime$ are the hopping amplitudes between the nearest-neighbour and next nearest-neighbor O $2p_x$ and $2p_y$ orbitals (light blue and dark blue arrows, respectively). $U_d$ is the onsite electron-electron repulsion on the Cu orbitals. The CDMFT method used in this work embeds a cluster of 12 lattice sites (solid green square) in a self-consistent bath of noninteracting electrons. 
(b, c) Sketch of the Cu $3d$ (solid red line) and O $2p$ (dashed blue line) partial density of states for a charge-transfer insulator and a Mott-Hubbard insulator. For the former, the Fermi level lies between the upper Hubbard band (UHB) and the charge-transfer band (CTB), so that $U_d > \Delta$. For the latter, the Fermi level lies between the lower (LHB) and the upper Hubbard band, so that $U_d < \Delta$. Here, $\Delta=\epsilon_d +U_d - \tilde{\epsilon}_p$ is the energy difference between the upper Hubbard band at $\epsilon_d +U_d$ and the onsite energy of the O $2p$ orbitals, where we renormalised $\epsilon_p$ by $\tilde{\epsilon}_p=\epsilon_p -2t_{pp}$. 
}
\label{fig:EmeryModel}
\end{figure}

The Emery model~\cite{Emery_1987} we study has the Hamiltonian:
\begin{align}
H & = \sum_{\mathbf{k} \sigma} C_{\mathbf{k} \sigma}^{\dagger}  \left( \mathbf{h}_{0}(\mathbf{k}) -\mu \mathbf{I} \right)  C_{\mathbf{k} \sigma} 
+ U_d \sum_{\mathbf{R}_{i}} n_{d \mathbf{R}_{i} \uparrow} n_{d \mathbf{R}_{i} \downarrow} . 
\label{eq:EmeryModel}
\end{align}
Here, in the vectors $C_{\mathbf{k} \sigma}^{\dagger} = \left( d_{\mathbf{k} \sigma}^{\dagger} ,  p_{x \mathbf{k}  \sigma}^{\dagger} , p_{y \mathbf{k} \sigma}^{\dagger}  \right)$ and $C_{\mathbf{k} \sigma} = \left( d_{\mathbf{k} \sigma},  p_{x \mathbf{k} \sigma}, p_{y \mathbf{k} \sigma} \right)^{T}$, $d_{\mathbf{k} \sigma}^{\dagger}$ and $d_{\mathbf{k} \sigma}$ creates and annihilates an electron with momentum $\bf{k}$ and spin $\sigma$ in the Cu $3d_{x^2-y^2}$ orbital and $p_{\alpha \mathbf{k}  \sigma}^{\dagger}, p_{\alpha \mathbf{k} \sigma}$ ($\alpha=x,y$) creates and annihilates an electron with momentum $\bf{k}$ and spin $\sigma$ in the O $2p_{\alpha}$ orbital, $n_{d \mathbf{R}_{i} \sigma} = d_{ \mathbf{R}_{i} \sigma}^{\dagger} d_{ \mathbf{R}_{i} \sigma}$ is the number operator at the Cu site $\mathbf{R}_{i}$, $\mu$ is the chemical potential and $U_d$ is the onsite Coulomb repulsion on the Cu orbital. 

In the matrix
\begin{align}
\mathbf{h}_0 (\mathbf{k}) & =
\left(
\begin{array}
[c]{ccc}
\epsilon_{d} & V_{dp_x} & V_{d p_y} \\
V^\dagger_{dp_x} & \tilde{\epsilon}_p + W_{p_x p_x} & W_{p_x p_y} \\
V^\dagger_{dp_y} & W^\dagger_{p_x p_y} & \tilde{\epsilon}_p + W_{p_y p_y} 
\end{array} \right)  , 
\label{eq:h0}
\end{align}
the parameters $\epsilon_{d}$ and $\epsilon_{p} = \epsilon_{p_x} = \epsilon_{p_y}$ are the onsite energy of the Cu $3d_{x^2-y^2}$ orbital and O $2p_{x}, 2p_{y}$ orbitals, $V_{dp_\alpha} = t_{pd}\left(  1-e^{-ik_{\alpha}}\right)$, $W_{p_\alpha p_\alpha} = 2t_{pp}^\prime \cos k_{\alpha}$, and $W_{p_x p_y} = t_{pp}\left(  1-e^{ik_{x}}\right)  \left(  1-e^{-ik_{y}}\right)$. 
Here, we have set the Cu-Cu lattice distance to unity, $t_{pd}$ is the hopping amplitude between nearest neighbor Cu-O orbitals, $t_{pp}$ and $t_{pp}^\prime$ are the hopping amplitudes between nearest neighbor and next nearest neighbor O-O orbitals. The phase factors of the hopping amplitudes are shown in Fig.~\ref{fig:EmeryModel}(a). Following Ref.~\cite{AndersenLDA}, in Eq.~\ref{eq:h0} we renormalised the onsite energy of the O orbitals by replacing $\epsilon_p$ by $\tilde{\epsilon}_p = \epsilon_p -2t_{pp}$. 

We study the Emery model Eq.~\ref{eq:EmeryModel} at finite temperature using the cellular extension~\cite{maier, kotliarRMP, tremblayR} of dynamical mean-field theory~\cite{rmp}. We embed a cluster of 12 lattice sites (solid green square in Fig.~\ref{fig:EmeryModel}a) containing $N_d=4$ Cu sites and $N_p=8$ O sites into a self-consistent bath of noninteracting electrons. We solve the resulting cluster plus bath model with the hybridization expansion continuous-time quantum Monte Carlo method (CT-HYB)~\cite{millisRMP, Werner:2006, hauleCTQMC, patrickSkipList} in the segment representation~\cite{millisRMP}. Further details about the method are described in Refs.~\cite{Lorenzo3band, Nicolas:PNAS2021, Nicolas:Master}. 

Using $t_{pp}=1$ as our unit of energy and temperature $T$ (with $k_B=1$), we study the set of parameters
\begin{align}
\epsilon_d =0, t_{pp}^\prime=1, t_{pd}=1.5 , T=1/20. 
\label{eq:SetParameters}
\end{align}
To tune the Emery model from the charge-transfer to the Mott-Hubbard regime, we vary the onsite O energy $\epsilon_p$ in the range $\epsilon_p \in [-4, 10]$ and the onsite Coulomb interaction $U_d$ in the range $U_d \in [2, 15]$. We change the chemical potential $\mu$ to study a doping range $\delta=5-n_{\rm tot} \in [-0.2, 0.2]$ (where $\delta>0$ indicates hole doping and $\delta<0$ indicates electron doping). 
 
To gain some insights into these sets of parameters, let us consider some limiting cases. For $U_d=0$ and $t_{pd}=0$, the parameters $\epsilon_d =0, \epsilon_p=10$ set the Cu $3d$ level below the O $2p$ bands. For $U > \tilde{\epsilon}_p -\epsilon_d$, the Cu $3d$ level is split into two levels of energy $\epsilon_d$ and $\epsilon_d +U_d$, with the latter lying above the O $2p$ bands. For a total occupation $n_{\rm tot} = n_d +2n_p = 5 $, the model shows a charge-transfer gap between the O $2p$ bands and the Cu $3d$ level at energy $\epsilon_d +U_d$. 
On the other hand, for $U_d=0$ and $t_{pd}=0$, the parameters $\epsilon_d =0, \epsilon_p=-4$ set the Cu $3d$ level above the O $2p$ bands. As before, the interaction strength $U_d$ splits the Cu $3d$ level into two levels at $\epsilon_d$ and $\epsilon_d +U_d$. For $n_{\rm tot}=5$, the model now shows a Mott-Hubbard gap between the Cu $3d$ levels. 
If we now turn on the Cu-O hopping $t_{pd}$, the Cu $3d$ levels evolves into bands and the Cu $3d$ and O $2p$ bands develop a mixed $d$-$p$ character. A sketch of the resulting local density of states is shown in Fig.~\ref{fig:EmeryModel}(b,c). 

With the same methodology used here, Ref.~\cite{Lorenzo3band} studied $\epsilon_p=9$ for hole doping and Ref.~\cite{GiovanniPRB2025} studied $\epsilon_p=8$ for either hole or electron doping. We extend Refs.~\cite{Lorenzo3band, GiovanniPRB2025} by studying several values of the onsite O energy $\epsilon_p$ in the range $\epsilon_p \in [-4, 10]$ and of the interaction strength $U_d$, for either hole or electron doping. Contrary to Refs.~\cite{Lorenzo3band, GiovanniPRB2025}, here we fix the temperature at $T=1/20$ since we cannot access lower temperatures due to the Monte Carlo sign problem that becomes more severe for small values of $\epsilon_p$ (see Appendix~\ref{appendix}).

\section{Zero doping}
\label{sec:zerodoping}

In this section we study the Emery model Eq.~\ref{eq:EmeryModel} at zero doping, where the CuO$_2$ unit cell has an odd number of electrons (five electrons, or equivalently one hole). We construct the Zaanen-Sawatzky-Allen diagram to map out the correlated insulating regimes induced by the onsite Coulomb repulsion on the Cu orbitals. Then we quantify the degree of sharing of the electrons among Cu and O orbitals and we examine the correlated gap size. 

\subsection{Zaanen-Sawatzky-Allen diagram}

\begin{figure}[ht!]
\centering{
\includegraphics[width=1.0\linewidth]{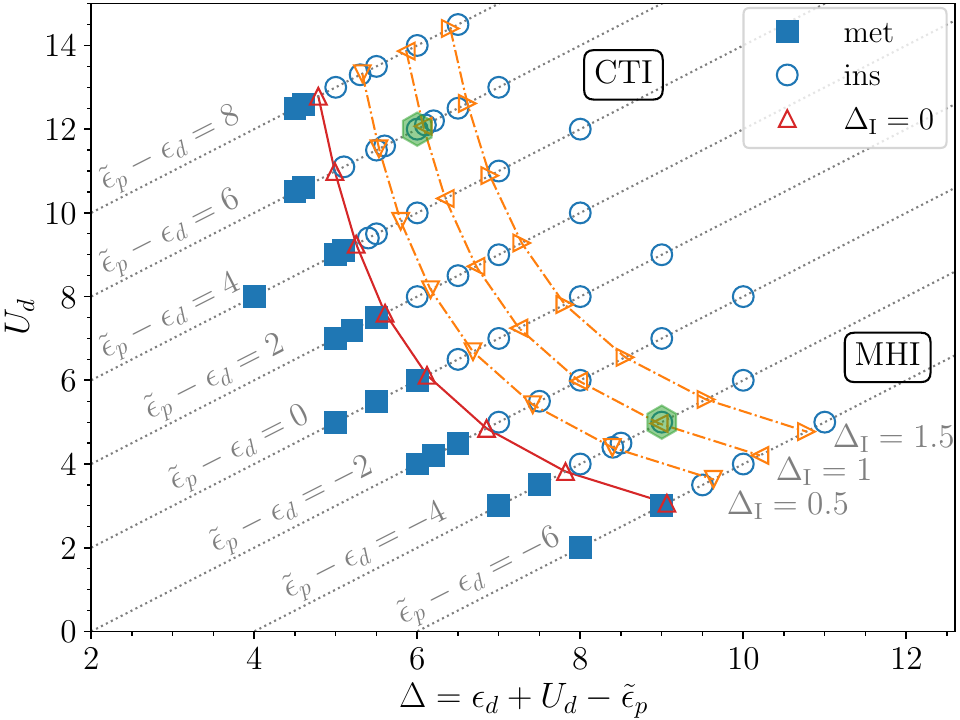}}
\caption{Computed Zaanen-Sawatzky-Allen diagram $U_d$ vs $\Delta = \epsilon_d +U_d - \tilde{\epsilon}_p$ for the Emery model at $n_{\rm tot} = 5$, i.e. at $\delta=0$. Dotted grey lines denote constant values of $\tilde{\epsilon}_p -\epsilon_d$. Filled blue squares indicate a metallic state. Open blue circles indicate an insulating state, of charge-transfer type (CTI) or Mott-Hubbard type (MHI). The metal to insulator boundary (red up triangles) is determined by the loci where the insulating charge gap closes, i.e. $\Delta_{\rm I}=0$ (see Fig.~\ref{fig:gap}a). Orange dot-dashed lines with down, left and right triangles indicate the lines of constant charge gap size $\Delta_{\rm I} = 0.5, 1, 1.5$ (see Fig.~\ref{fig:gap}a).  
}
\label{fig:zsa}
\end{figure}

The Emery model can describe different correlated insulators taking place for $n_{\rm tot} = n_d + 2n_p = 5$, or equivalently one hole per CuO$_2$ unit cell. A useful way to show this is to calculate the Zaanen-Sawatzky-Allen diagram~\cite{zsa}. Note that in our convention, $n_d$ and $n_p$ vary between 0 and 2, where 0 denotes an empty orbital and 2 denotes a full orbital. 

Fig.~\ref{fig:zsa} shows the calculated Zaanen-Sawatzky-Allen diagram at $T=1/20$ with CDMFT. On the $y$-axis, there is the onsite interaction strength $U_d$ on the Cu orbital. On the $x$-axis, there is the bare charge-transfer energy $\Delta \equiv \epsilon_d +U_d - \tilde{\epsilon}_p$ in the atomic limit. To map out the Zaanen-Sawatzky-Allen diagram, we perform scans along the diagonal lines (gray dotted lines), which correspond to different values of the bare Cu-O energy distance $\tilde{\epsilon}_p - \epsilon_d$, ranging from $-6$ to $8$ at a step of $2$. Ref.~\cite{Lorenzo3band} studied $\tilde{\epsilon}_p -\epsilon_d = 7$, and Ref.~\cite{GiovanniPRB2025} explored $\tilde{\epsilon}_p -\epsilon_d = 6$ with the same model and method used here. The normal state of the Emery model shows two main phases, one metallic phase (filled blue squares) and one correlated insulating phase (open blue circles). We identify these phases by calculating the total occupancy $n_{\rm tot}$ versus the chemical potential $\mu$. In a correlated insulator, for fixed $\tilde{\epsilon}_p - \epsilon_d$ and $T$, $n_{\rm tot}$ vs $\mu$ shows a plateau at $n_{\rm tot} = 5$. Therefore, if $dn_{\rm tot}/d\mu$ is smaller than an arbitrary small value (here $0.005$, as in Ref.~\cite{GiovanniPRB2025}), we characterise the system as a correlated insulator. Otherwise, we identify the system as metallic.

In the Zaanen-Sawatzky-Allen diagram $U_d-\Delta$, the boundary between the metal and correlated insulator (solid red line with up triangles) decreases monotonically with increasing $\Delta$. In Ref.~\cite{Lorenzo3band} it was shown for $\tilde{\epsilon}_p -\epsilon_d = 7$ that this boundary at low temperatures is first order in character. Note also that the location of this boundary depends on the short-range correlations included in our method. Neglecting such correlations would increase the critical value of $U_d$ for opening a correlated insulator, as shown for example in the context of the single-band Hubbard model~\cite{phk} and of the $d$-$p$ model~\cite{go}. 
Physically, we can have a correlated insulator for large values of $U_d$ and moderate values of $\Delta$, or for small values of $U_d$ and large values of $\Delta$. This corresponds to two different types of correlated insulators~\cite{zsa}: charge-transfer insulator (CTI) for $U_d \gtrsim \Delta$ and Mott Hubbard insulator (MHI), for $U_d \lesssim \Delta$. There is a crossover between these two regimes, which is known as the intermediate regime, for $U_d \approx \Delta$.

\begin{figure}%[ht!]
\centering{
\includegraphics[width=1.0\linewidth]{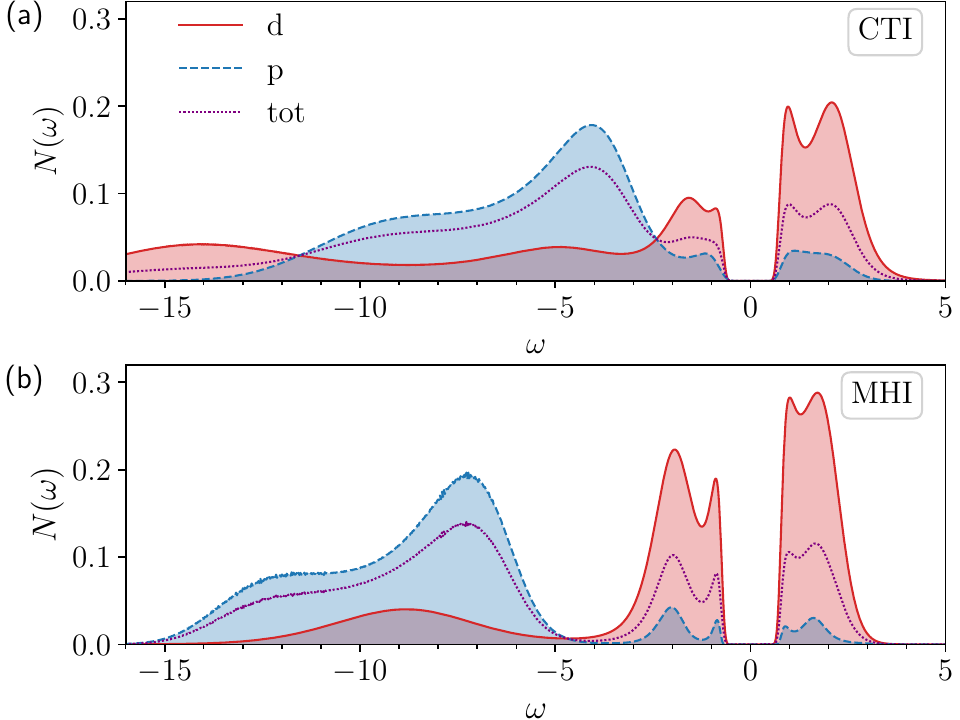}}
\caption{Partial density of states $N(\omega)$ of the Cu $3d$ orbital (red solid line) and of the degenerate O $2p_x$, $2p_y$ orbitals (blue dashed line), along with the total density of states $N_{\rm tot} (\omega) = (N_d(\omega) + 2N_p(\omega))/3$ (purple dotted line), in the charge-transfer insulator (a) and in the Mott Hubbard insulator (b). Data are for the set of parameters of Eq.~\ref{eq:SetParameters} and $U=12.1, \tilde{\epsilon}_p -\epsilon_d = 6$ (panel (a)) and $U=5, \tilde{\epsilon}_p -\epsilon_d=-4$ (panel (b)), and correspond to the green hexagons in Fig.~\ref{fig:zsa}. 
}
\label{fig:DOS-ins}
\end{figure}

To further characterise the charge-transfer and Mott-Hubbard insulators, Fig.~\ref{fig:DOS-ins} shows the projected local density of states on copper $3d$ (red) and oxygen $2p$ (blue) orbitals, as well as the total density of states (purple). Data are shown for a charge-transfer insulator and a Mott-Hubbard insulator, corresponding to the values of $U_d$ and $\Delta$ marked by the green hexagons in the Zaanen-Sawatzky-Allen diagram of Fig.~\ref{fig:zsa}. We perform the analytical continuation from Matsubara to real frequencies using the method of Ref.~\cite{DominicMEM}. 
In the charge-transfer insulator (Fig.~\ref{fig:DOS-ins}(a)), there is an insulating gap between the charge-transfer band and the upper Hubbard band. In contrast, in the Mott-Hubbard insulator (Fig.~\ref{fig:DOS-ins}(b)) there is an insulating gap between the lower and upper Hubbard band. They both have a mixed $d$-$p$ character, which is more pronounced in the charge-transfer regime than the Mott-Hubbard regime. The next subsection quantifies the degree of the mixed $d$-$p$ character of these states.

\subsection{Degree of charge shared among Cu and O orbitals}
\label{sec:sharing}
\begin{figure}[ht!]
\centering{
\includegraphics[width=1.0\linewidth]{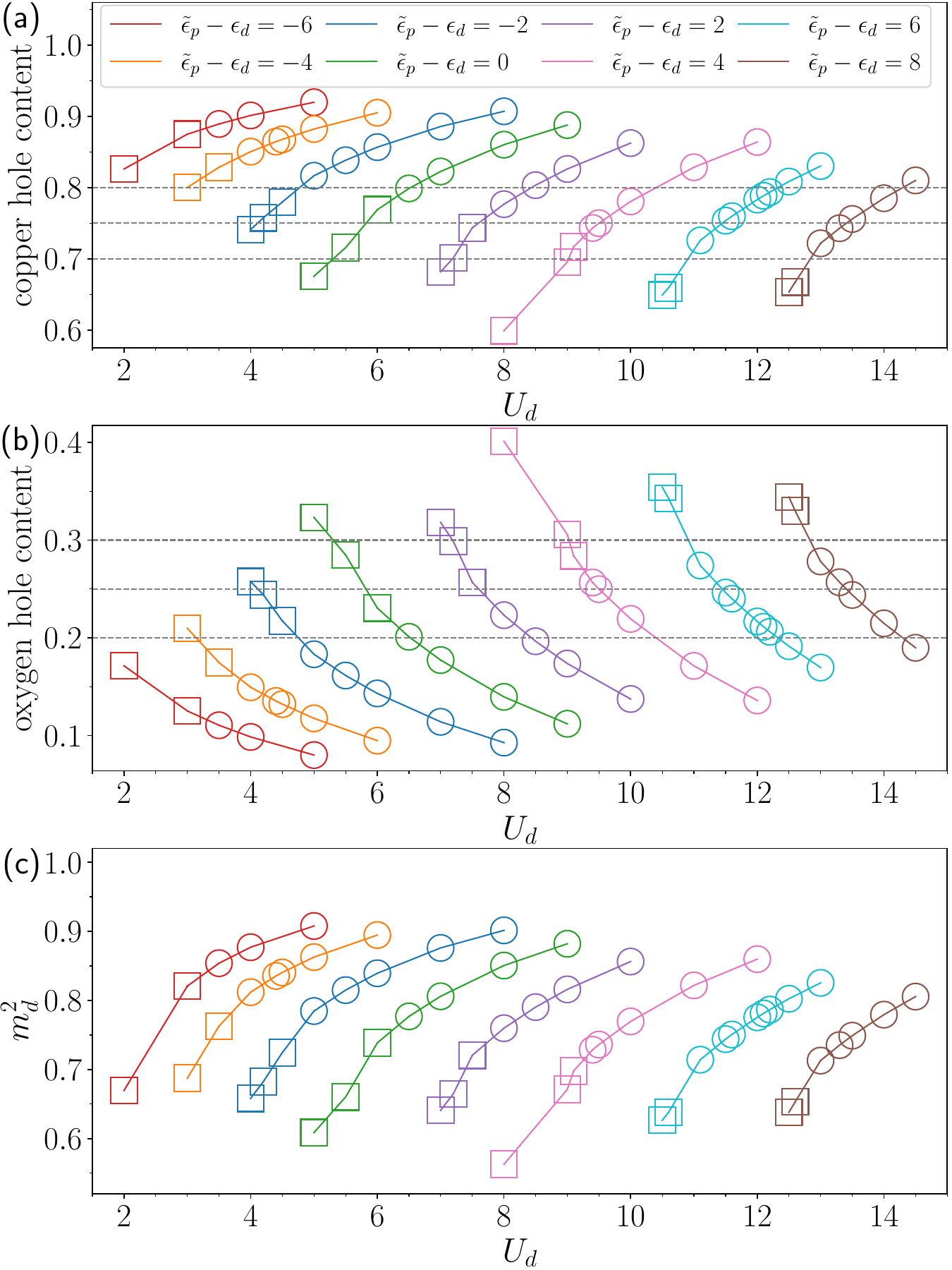}
}
\caption{(a) Cu hole content (defined as $p_d = 2-n_d)$), (b) O hole content (defined as $p_p = 2(2-n_p)$), and (c) Cu local moment squared (defined as $m_d^2 =n_d -2D$) as a function of $U_d$. Data are taken at $\delta=0$, for different values of the bare Cu-O energy difference $\tilde{\epsilon}_p -\epsilon_d$, and for the set of parameters of Eq.~\ref{eq:SetParameters}. Circles (squares) denote insulating (metallic) states, as in Fig.~\ref{fig:zsa}. The intersection of the curves in panels (a) and (b) with horizontal gray dashed lines defines the loci of constant Cu hole content $p_d=0.7, 0.75, 0.8$ (O hole content $p_p=0.3, 0.25, 0.2$, respectively). These loci are indicated by cyan dashed lines in the Zaanen-Sawatzky-Allen diagram of Fig.~\ref{fig:discussion}a. 
}
\label{fig:zero-doping}
\end{figure}

The Zaanen-Sawatzky-Allen diagram characterises the Emery model at $n_{\rm tot} = n_d +2n_p = 5$, or equivalently $\delta =0$, or one hole per CuO$_2$ unit cell. The hybridization $t_{pd}$ leads to a mixed character of this hole. As a consequence, the Cu orbital is slightly more than half filled and the O orbital is almost fully filled, i.e. $n_d = 1 +2\eta$ and correspondingly $n_p = 2 -\eta$. In the hole representation, this leads to define the copper hole content as $p_d = 2-n_d$ and the oxygen hole content as $p_p = 2(2-n_p)$, where the extra factor of 2 takes into account the $2p_x, 2p_y$ orbital degeneracy. Hence, the parameter $\eta$ can be considered as a measure of the mixed $d$-$p$ character of the hole. This section studies the orbital character of the hole, i.e. how this hole is shared among Cu and O orbitals in the regions described by the Zaanen-Sawatzky-Allen diagram.

Figures~\ref{fig:zero-doping}(a,b) show the copper hole content and oxygen hole content as a function of $U_d$ for different values of $\tilde{\epsilon}_p -\epsilon_d$. The mixed character of the hole decreases with increasing $U_d$. Therefore, upon increasing $U_d$ at fixed $\tilde{\epsilon}_p -\epsilon_d$, the copper hole content increases (i.e. $n_d \rightarrow 1$ for $U_d \rightarrow \infty$) and the oxygen hole content decreases (i.e. $n_p \rightarrow 2$ for $U_d \rightarrow \infty$). 
Conversely, on increasing $\tilde{\epsilon}_p -\epsilon_d$ at fixed $U_d$, the mixed character of the hole increases. Therefore, on increasing $\tilde{\epsilon}_p -\epsilon_d$ at fixed $U_d$, the copper hole content decreases (i.e. $n_d \rightarrow 1$ for $\epsilon_p \rightarrow -\infty$) and the oxygen hole content increases (i.e. $n_p \rightarrow 2$ for $\epsilon_p \rightarrow -\infty$). 

Since the Cu orbital is slightly more than half filled (i.e. $n_d = 1+2\eta$), it is instructive to analyse its associated spin local moment. Figure~\ref{fig:zero-doping}(c) shows the local moment squared on the Cu orbital, $m_d^2 =  (n_{d \uparrow} - n_{d \downarrow})^2  = n_d -2D$, where $D$ is the double occupancy, as a function of $U_d$ for different values of $\tilde{\epsilon}_p -\epsilon_d$. 
Upon increasing $U_d$ at fixed $\tilde{\epsilon}_p -\epsilon_d$, $m_d^2$ increases, as the double occupancy of the Cu orbitals becomes energetically unfavourable. Upon increasing $\tilde{\epsilon}_p -\epsilon_d$ at fixed $U_d$, $m_d^2$ decreases due to the increased mixed character of the hole (i.e. $n_d \rightarrow 1$ for $\epsilon_p \rightarrow -\infty$). 

At lower temperatures than that used here, it is expected that the large local moments on nearest Cu orbitals may arrange themselves in an antiferromagnetic pattern~\cite{Scalettar:PRB1991, ArrigoniCuO2, Cui:PRR2020}, owing to the superexchange mechanism~\cite{AndersonSE, Anderson:1987, Dagotto:RMP1994}. Note however, that thermal fluctuations prevent long-range order in two dimensions~\cite{MWtheorem}, but in real materials, small inter-plane coupling suffice to induce long-range order at a small but non-zero temperature.

\subsection{Correlated charge gap}

\begin{figure}[ht!]
\centering{
\includegraphics[width=1.0\linewidth]{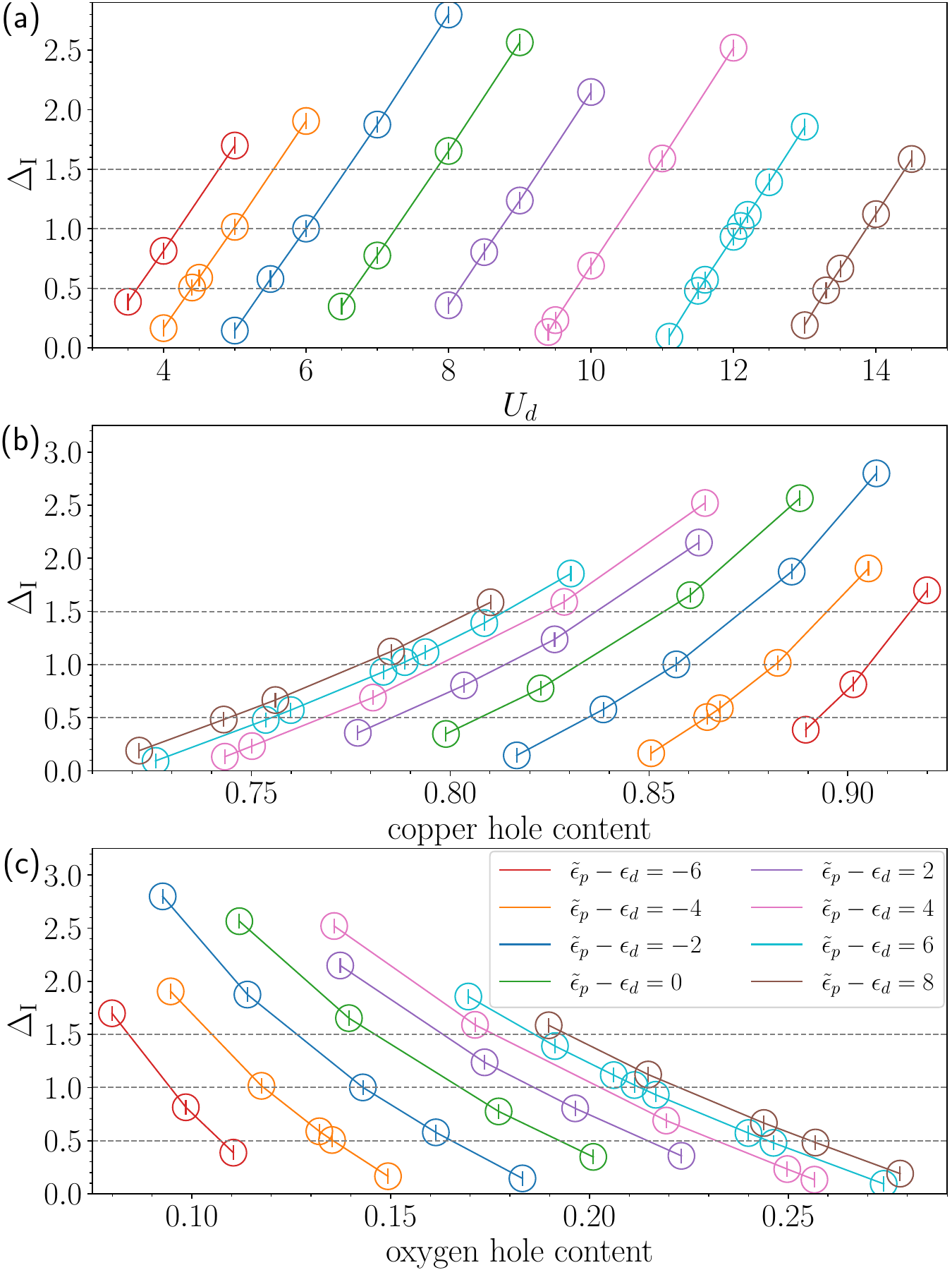}}
\caption{(a) Insulating charge gap size $\Delta_{\rm I}$ vs interaction strength $U_d$ for different values of $\tilde{\epsilon}_p -\epsilon_d$. The intersection of the linear fit of the curves $\Delta_{\rm I}(U_d)$ with the $x$ axis, i.e. $\Delta_{\rm I} (U_d)=0$, determines the metal to insulator boundary in the Zaanen-Sawatzky-Allen diagram of Fig.~\ref{fig:zsa}. The intersection of the linear fit of the data with horizontal gray dashed lines defines the loci at constant charge gap size $\Delta_{\rm I} =0.5, 1, 1.5$ in the Zaanen-Sawatzky-Allen diagram of Fig.~\ref{fig:zsa}. (b) Insulating charge gap size $\Delta_{\rm I}$ vs copper hole content for different values of $\tilde{\epsilon}_p -\epsilon_d$. (c) Insulating charge gap size $\Delta_{\rm I}$ vs oxygen hole content for different values of $\tilde{\epsilon}_p -\epsilon_d$. Other model parameters are as in Eq.~\ref{eq:SetParameters}.
}
\label{fig:gap}
\end{figure}

Next we study the size of the charge gap of the correlated insulators, $\Delta_{\rm I}$. We determine $\Delta_{\rm I}$ by the width of the plateau at $n_{\rm tot}(\mu) =5$ (numerically we take $dn_{\rm tot}/d\mu < 0.005$). The quantity $\Delta_{\rm I}$ is comparable to the spectral gap in the local density of states (see Fig.~\ref{fig:DOS-ins}). However, $\Delta_{\rm I}$ has the advantage of not relying on the analytical continuation.  

Figure~\ref{fig:gap}(a) shows the charge gap size $\Delta_{\rm I}$ versus the interaction strength $U_d$ for different values of $\tilde{\epsilon}_p -\epsilon_d$. As expected, the size of the gap increases with increasing $U_d$. The critical value of the interaction strength for opening a correlated insulator, $U_{d, \rm MIT}$, is given by the $x$-axis intercept of the linear fit of the curves $\Delta_{\rm I}(U_d)$ and monotonically increases with increasing $\tilde{\epsilon}_p -\epsilon_d$. The resulting metal-insulator boundary $U_{d, \rm MIT}(\Delta)$ is shown by the red line with up triangles in the Zaanen-Sawatzky-Allen diagram of Figure~\ref{fig:zsa}. Note that at smaller temperatures than that considered here, the metal to insulator boundary at $n_{\rm tot} =5$ is slightly shifted and becomes first-order, as shown in Ref.~\cite{Lorenzo3band}. 
Similarly, the $x$-intercepts of the curves with horizontal lines determines the lines at constant charge gap size and are shown with orange dot-dashed lines in Figure~\ref{fig:zsa} for $\Delta_{\rm I} = 0.5, 1, 1.5$. As shown in in Fig.~\ref{fig:zsa}, these isolines of charge gap size are approximately parallel to the metal-insulator boundary. 

Next we study the evolution of the charge gap size as a function of Cu or O hole content, as shown Figure~\ref{fig:gap}(b,c). On increasing the copper hole content at fixed $\tilde{\epsilon}_p -\epsilon_d$, the gap size $\Delta_{\rm I}$ increases. Conversely, on increasing $\tilde{\epsilon}_p -\epsilon_d$ at fixed gap size, the copper hole content decreases. 
Since at $\delta=0$ the O hole content and Cu hole content are related by $p_p = 1-p_d$, opposite trends occur for the O hole content: upon increasing the oxygen hole content at fixed $\tilde{\epsilon}_p -\epsilon_d$, the charge gap size $\Delta_{\rm I}$ decreases. Conversely, on increasing $\tilde{\epsilon}_p -\epsilon_d$ at fixed gap size, the oxygen hole content increases.

Refs.~\cite{Nicolas:PNAS2021, Davis:PNAS2022, Jurkutat:PNAS2023} pointed out that the gap size of the charge-transfer insulator is anticorrelated with the O hole content. Our findings confirm that this result is valid not only for the charge-transfer regime, but all the way from the charge-transfer to the Mott-Hubbard regime as well. 

In addition, our results show that the absolute value of the slope of the charge gap size as a function of either copper hole content or oxygen hole content is reduced at small charge gap size (Fig.~\ref{fig:gap}(b,c)). This is consistent with the results of Ref.~\cite{go}, which ascribes the reduced slope of $\Delta_{\rm I}$ to the increased short-range correlations as the metal to insulator boundary is approached.

\section{Finite doping}
\label{sec:finitedoping}

Having analysed the zero doping case, we now turn to study the effect of modifying the number of charge carriers in the CuO$_2$ plane by either electron or hole doping. We show how doping induces a transfer of spectral weight in the local density of states and then we quantify the charge redistribution among Cu and O orbitals induced by doping using the partial occupancies of the orbitals.

\subsection{Local density of states}

\begin{figure}[ht!]
\centering{
\includegraphics[width=1.0\linewidth]{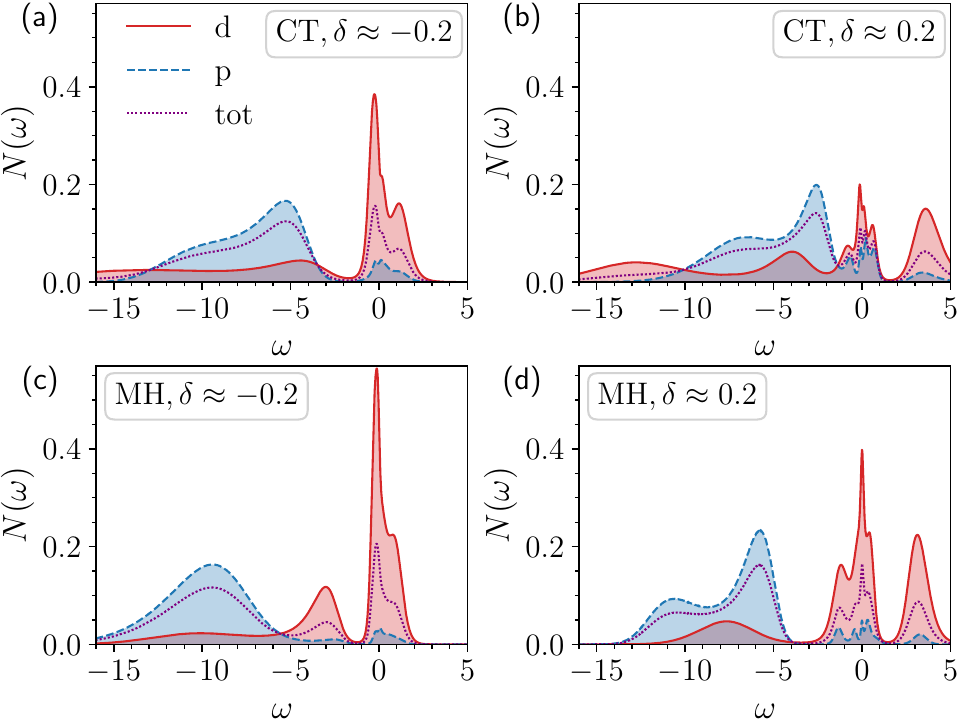}}
\caption{Partial density of states $N(\omega)$ of the Cu $3d$ orbital (red solid line) and of the O $2p_x$, $2p_y$ orbitals (blue dashed line), along with the total density of states $N_{\rm tot} (\omega) = (N_d(\omega) + 2N_p(\omega))/3$ (purple dotted line), for $20\%$ electron and hole doping of the charge-transfer regime (panel (a) and (b), respectively), and of the Mott-Hubbard regime (panel (c) and (d), respectively). Data are for the set of parameters of Eq.~\ref{eq:SetParameters} and $U=12.1, \tilde{\epsilon}_p -\epsilon_d = 6$ (panels (a,b)) and $U=5, \tilde{\epsilon}_p -\epsilon_d =-4$ (panels (c,d)).  
}
\label{fig:DOS-doping}
\end{figure}

Let us consider the effect of adding or removing electrons in the density of states of the charge-transfer and Mott-Hubbard insulating states of Fig.~\ref{fig:DOS-ins}. 
Figs.~\ref{fig:DOS-doping}(a),(b) show the projected local density of states on copper 3d (red) and oxygen 2p (blue) orbitals, as well as the total density of states (purple) at doping values $\delta=\pm 0.2$, for a doped charge-transfer insulator (panels (a), (b)) and a doped Mott-Hubbard insulator (panels (c), (d)). 

Let us first consider the doped charge-transfer insulators in Fig.~\ref{fig:DOS-doping}(a,b). On electron doping (Fig.~\ref{fig:DOS-doping}(a)), there is a transfer of spectral weight from high to low energies. This process introduces states into the upper Hubbard band and leads to a peak in the density of states close to the Fermi level at $\omega=0$. The peak has a predominant $d$ character. On hole doping, there is a transfer of spectral weight into the charge-transfer band, which produces a peak at the Fermi level with a large mixed $d$-$p$ character. 
Let us now turn to the doped Mott-Hubbard insulators in Fig.~\ref{fig:DOS-doping}(c,d). On electron (hole) doping, the dopant carriers enter the upper Hubbard band (lower Hubbard band), giving rise to a metallic state characterised by a peak close to $\omega=0$ of mainly $d$ character. 

Comparing the local density of states in Fig.~\ref{fig:DOS-doping}(a,b) and Fig.~\ref{fig:DOS-doping}(c,d), there is an asymmetry about which orbital the dopant carriers go upon electron or hole doping. This asymmetry decreases as the system evolves from the charge-transfer regime to the Mott-Hubbard regime. 
In the former, the doped holes introduces states in a charge-transfer band with mainly $p$ character and doped electrons introduces states in the upper Hubbard band with mainly $d$ character. In the latter, either doped electrons or holes introduces states in bands of mainly $d$ characters, the Hubbard bands. 

Finally, note that our analysis of the density of states is restricted at $\delta=\pm 0.2$ and $T=1/20$. Ref.~\cite{GiovanniPRB2025} shows, for $\tilde{\epsilon}_p -\epsilon_d = 6$ and smaller temperatures, that the metallic state found upon small either electron or hole doping a charge-transfer insulator has the features of a strongly correlated pseudogap~\cite{kyung, GiovanniPRB2025}. We leave as an interesting future work how the pseudogap evolves with the bare Cu-O energy distance.

\subsection{Charge redistribution among Cu and O orbitals induced by doping}

\begin{figure*}%[ht!]
\centering{
\includegraphics[width=0.99\linewidth]{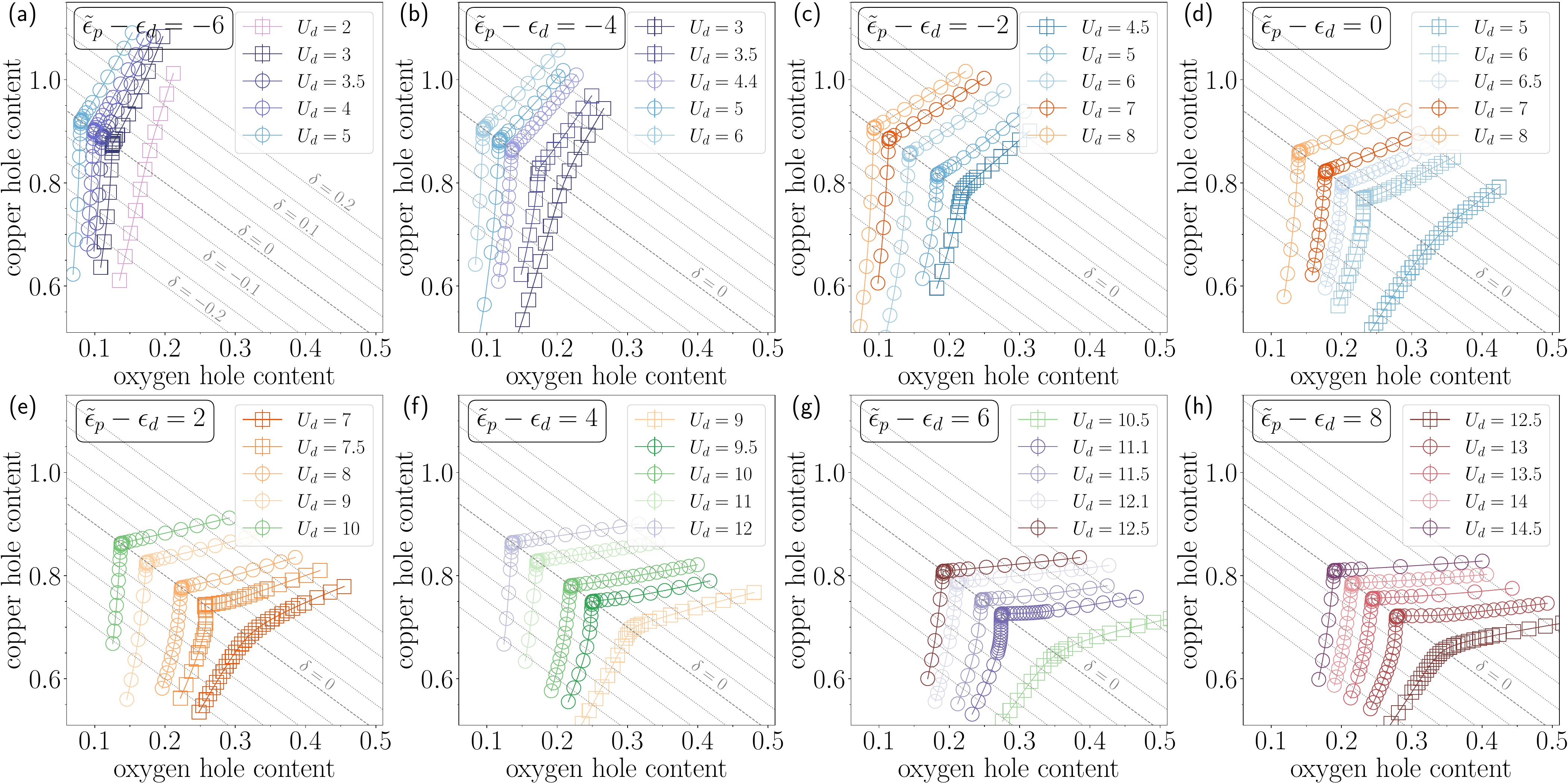}}
\caption{Cu hole content vs O hole content for different values of $U_d$. Each panel corresponds to a fixed value of $\tilde{\epsilon}_p -\epsilon_d$. Lines with circles (squares) denote that the system at $\delta=0$ is a correlated insulator (metal). In each panel, the grey dashed line indicates $\delta=0$. The grey dotted lines mark constant values of electron or hole doping, as indicated in panel (a). Other model parameters are as in Eq.~\ref{eq:SetParameters}.
}
\label{fig:choVSohc}
\end{figure*}

Following Refs.~\cite{Jurkutat:PRB2014, Rybicki:NatComm2016, Jurkutat:PNAS2023}, a compact way to quantify the effect of the electron or hole doping on the charge redistribution among Cu and O orbitals is to plot the Cu hole content, i.e. $2 -n_d$, vs O hole content, i.e. $2(2 -n_p)$. 
This is shown in Fig.~\ref{fig:choVSohc} for different values of $\tilde{\epsilon}_p -\epsilon_d$. Each panel contains different values of correlation strength $U_d$, corresponding to selected values of $U_d$ in Fig.~\ref{fig:zsa}. 
The zero doping state analysed in Sec.~\ref{sec:zerodoping} corresponds to the dashed grey diagonal line in each panel of Fig.~\ref{fig:choVSohc}. Parallel dotted grey lines indicate constant values of electron doping (bottom left region) or hole doping (top right region).

Above $U_{d, \rm MIT}$ (curves with circles) and for electron doping, the curves have large positive slopes: electron doping mostly decreases the Cu hole content whilst having small effect on the O hole content, all the way from the Mott-Hubbard to the charge-transfer regime (i.e. upon increasing $\tilde{\epsilon}_p -\epsilon_d$, from panel (a) to (h)). 
Conversely, hole doping is markedly different between Mott-Hubbard and charge-transfer regimes: above $U_{d, \rm MIT}$ and for hole doping, the slope of the curves decreases with increasing $\tilde{\epsilon}_p -\epsilon_d$, i.e. from the Mott-Hubbard to the charge-transfer regime. The almost flat slopes found in the charge-transfer regime indicates that hole doping mostly increases the O hole content whilst having a small effect on the Cu hole content. Physically, this reflects that doped holes mostly enter the O orbital in the charge-transfer regime. As the system evolves from the charge-transfer to the Mott-Hubbard regime, the doped holes find favourable to enter the Cu orbital. 

For a fixed value of $\tilde{\epsilon}_p -\epsilon_d$ and increasing $U_d$, we observe an overall shift of the curves towards the top left. Hence, the conclusion of Sec.~\ref{sec:sharing} extends to any doping level: the effect of increasing $U_d$ at any doping level is to reduce the oxygen hole content and to increase the copper hole content. This effect was already studied in Ref.~\cite{GiovanniPRB2025} for $\tilde{\epsilon}_p -\epsilon_d = 6$. 
For a fixed value of $U_d$ and increasing $\tilde{\epsilon}_p -\epsilon_d$, we observe an overall shift of the curves downwards towards the bottom right. This means that the effect of increasing $\tilde{\epsilon}_p -\epsilon_d$ is opposite to that of $U_d$, i.e. is to increase the oxygen hole content, and to decrease the copper hole content. 

Note that in Fig.~\ref{fig:choVSohc} we fix the temperature at $T=1/20$. Ref.~\cite{GiovanniPRB2025} shows, for $\tilde{\epsilon}_p -\epsilon_d=6$, that the effect of the temperature on the magnitude of the partial occupancies is small. Hence, we expect that the trends in Fig.~\ref{fig:choVSohc} are robust against the temperature. 

The trends emerging from Fig.~\ref{fig:choVSohc} give an effective framework for discussing the NMR experiments~\cite{Jurkutat:PRB2014, Rybicki:NatComm2016, Jurkutat:PNAS2023} on the charge redistribution among Cu and O orbitals of different cuprates. 
First, hole-doped cuprates are charge-transfer insulators, hence the relevant model parameters are $U_d > \Delta$ and $U_d > U_{d, {\rm MIT}}$, i.e. lines with circles in panels (e)-(h) of Fig.~\ref{fig:choVSohc}. 
Second, Refs.~\cite{Jurkutat:PRB2014, Rybicki:NatComm2016, Jurkutat:PNAS2023} show that the parent ($\delta=0$) compound of cuprates has a ratio of Cu/O hole content spanning from nearly 0.82/0.18 of La$_{2-x}$Sr$_x$CuO$_4$ (LSCO) to nearly 0.7/0.3 of YBa$_2$Cu$_3$O$_{6+y}$ (YBCO), to 0.5/0.5 of a Tl-based material. Therefore, our set of parameters seems to encompass the trends of LSCO and YBCO families only. 

\begin{figure*}%[ht!]
\centering{
\includegraphics[width=0.99\linewidth]{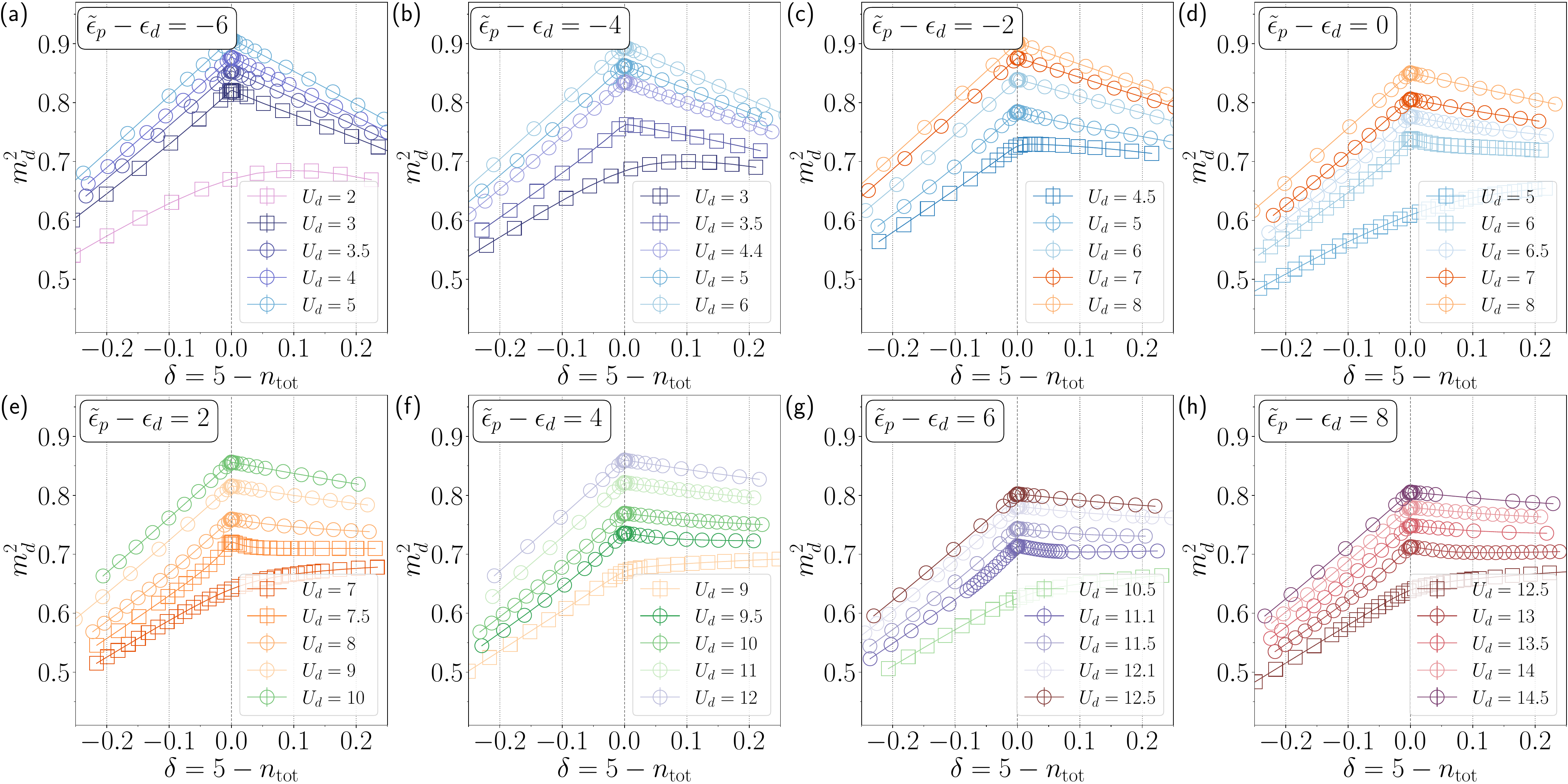}}
\caption{Local moment squared $m_d^2$ on the Cu orbital vs doping $\delta$. Each panel corresponds to a fixed value of the bare Cu-O energy distance $\tilde{\epsilon}_p -\epsilon_d$. Lines with circles (squares) denote that the system at $\delta=0$ is a correlated insulator (metal). In each panel, the grey dashed line indicates $\delta=0$. The grey dotted lines mark constant values of electron or hole doping. Other model parameters are as in Eq.~\ref{eq:SetParameters}.
}
\label{fig:moment}
\end{figure*}

Next, let us analyse the effect of doping on the local spin moment on the Cu orbitals. Fig.~\ref{fig:moment} shows the local spin moment squared on the Cu orbitals, $m_d^2=n_d -2D$, as a function of doping for different values of $\tilde{\epsilon}_p -\epsilon_d$ and $U_d$. 
Above $U_{d, \rm MIT}$ (curves with circles), doping suppresses the local moment. Depending on the values of $\tilde{\epsilon}_p -\epsilon_d$, this suppression can be asymmetrical upon electron or hole doping. 
In the Mott-Hubbard regime, doped holes mainly enter the Cu orbital, which increases $n_d$ and thus $D$, and therefore the asymmetry in the rate of decrease of $m_d^2(\delta)$ with increasing either electron or hole doping is less pronounced. 
In the charge-transfer regime, doped holes mainly enter the O orbitals. As a result, $m_d^2(\delta)$ decreases more gently on hole doping than on electron doping. 

Upon increasing $U_d$ at fixed $\tilde{\epsilon}_p -\epsilon_d$, the curves overall shift upward: the local moment squared on the Cu orbitals increases, since the double occupancy becomes less energetically favoured. 
Conversely, upon increasing $\tilde{\epsilon}_p -\epsilon_d$ at fixed $U_d$, the curves overall shift downward: the local moment squared decreases, due to the increased mixed character on the hole (i.e. $n_d \rightarrow 1$ for $\epsilon_p \rightarrow -\infty$). 

Finally, at lower temperatures than that explored here, studies on the Emery model have shown that both spin and charge may arrange themselves in spatially modulated patterns, such as spin or charge density waves~\cite{White:PRB2015, Kung:PRB2016, Ponsioen:PRB2023, Mai:PNAS2024}.

\section{Discussion}
\label{sec:discussion}

In Sections~\ref{sec:zerodoping} and \ref{sec:finitedoping} we quantified the variation of the charge redistribution among Cu and O orbitals in the normal state of the Emery model as a function of (i) the interaction strength $U_d$, (ii) the bare Cu-O energy difference $\tilde{\epsilon}_p-\epsilon_d$ and (iii) the doping. A key idea of our work is that further insights can be gained by discussing these results within the Zaanen-Sawatzky-Allen diagram. 

\subsection{Framing our results in the Zaanen-Sawatzky-Allen diagram}
\label{sec:discussion-theory}

\begin{figure*}%[ht!]
\centering{
\includegraphics[width=1.0\linewidth]{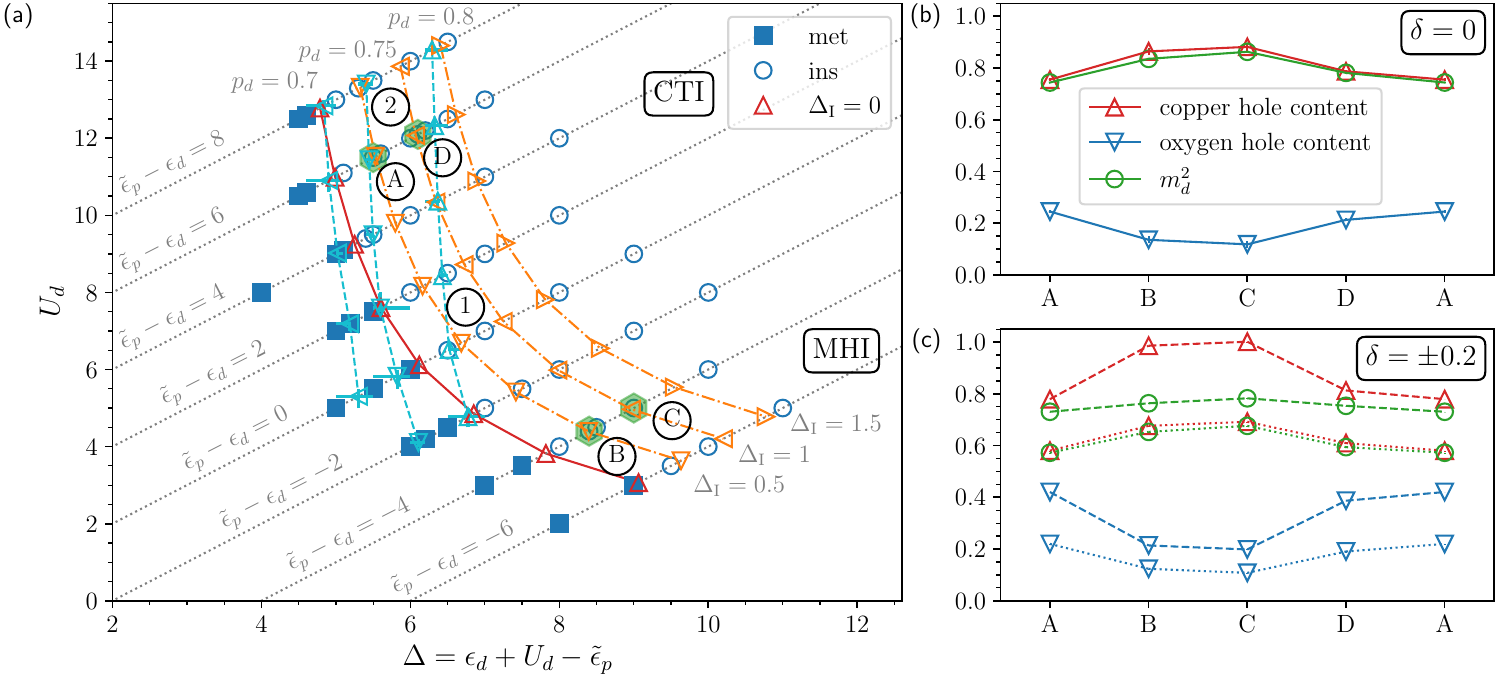}}
\caption{(a) Computed Zaanen-Sawatzky-Allen phase diagram $U_d - \Delta$ at $n_{\rm tot}=5$, i.e. at $\delta=0$. Data are the same as in Fig.~\ref{fig:zsa}. In addition to curves at constant charge gap size $\Delta_{\rm I} = 0.5, 1, 1.5$ (orange dot-dashed lines with down, left and right triangles), curves at constant Cu hole content $p_d=0.7, 0.75, 0.8$ are now added (cyan dashed lines with left, down and up triangles). The latter are determined in Fig.~\ref{fig:zero-doping}a, and correspond to curves at constant O hole content $p_p=0.3, 0.25, 0.2$. The charge redistribution among Cu and O orbitals along the path joining the points $A, B, C, D$ (green hexagons) is analysed in panels (b) and (c). The $(U_d, \tilde{\epsilon}_p -\epsilon_d)$ coordinates of the points $A, B, C, D$ are given by $A=(11.5, 6)$, $B=(4.4, -4)$, $C=(5,-4)$, $D=(12.1,6)$. Points 1 and 2 denote the intercepts between the curve at constant charge gap size $\Delta_{\rm I}=0.5$ and curves at Cu hole contents $0.8$ and $0.75$ (i.e. O hole contents $0.2$ and $0.25$), respectively. (b) Cu hole content (upper red triangles), O hole content (down blue triangles) and Cu local moment squared $m_d^2$ (green circles) evaluated at the points $A, B, C, D$ of panel (a), i.e. at $\delta=0$. (c) Same as panel (b), but Cu hole content, O hole content and Cu local moment squared are evaluated at the hole doping value $\delta = 0.20$ (dashed lines) and at the electron doping value $\delta=-0.20$ (dotted lines), relative to the points $A, B, C, D$ at $\delta=0$.
}
\label{fig:discussion}
\end{figure*}

To discuss our findings in the Zaanen-Sawatzky-Allen framework, let us consider the path $ABCD$ in the computed $U_d - \Delta$ Zaanen-Sawatzky-Allen phase diagram of Fig.~\ref{fig:discussion}(a). 
Fig.~\ref{fig:discussion}(b) shows the copper hole content (red up triangles), oxygen hole content (blue down triangles) and Cu local moment squared (green circles) evaluated at the points $A, B, C, D$, corresponding to the green hexagons in Fig.~\ref{fig:discussion}(a). 

Let us follow the path $A \rightarrow B$ that connects a charge-transfer insulator to a Mott-Hubbard insulator by reducing $\tilde{\epsilon}_p -\epsilon_d$ (i.e. increasing $\Delta$) and keeping constant the size of the insulating charge gap at $\Delta_{\rm I} \approx 0.5$. From $A$ to $B$, the copper hole content increases from $\approx 0.755$ to $\approx 0.865$ and correspondingly the oxygen hole content decreases from $\approx 0.245$ to $\approx 0.135$. 
Equivalently, this means that from $A$ to $B$ the {\it electronic} charge is transferred from Cu to O orbitals. Physically, from the charge-transfer to the Mott-Hubbard regime, the mixed $d$-$p$ character is reduced by suppressing the electronic charge on the Cu orbitals and redistributing it on the O orbitals. 
The spin Cu local moment squared $m_d^2$ follows the behavior of the Cu hole content. 
Physically, from the charge-transfer to the Mott-Hubbard regime, the electronic charge occupancy of the Cu orbital approaches unity ($n_d \rightarrow 1$) and thus $m_d^2$ is enhanced. 

Let us now follow the path $A \rightarrow D$, in which we start from a charge-transfer insulator and, by keeping $\tilde{\epsilon}_p -\epsilon_d$ fixed, we increase $U_d$ in such a way to double the size of the charge gap (from $\Delta_{\rm I} =0.5$ to $1.0$). From $A$ to $D$, the Cu hole content increases from $\approx 0.755$ to $\approx 0.787$ and correspondingly the O hole content decreases from $\approx 0.245$ to $\approx 0.213$. 
The Cu moment squared $m_d^2$ follows the behavior of the Cu hole content. 
Next we follow the path $B \rightarrow C$, in which we start from a Mott-Hubbard insulator and, by keeping $\tilde{\epsilon}_p -\epsilon_d$ fixed, we increase $U_d$ in such a way to double the size of the charge gap (from $\Delta_{\rm I} = 0.5$ to $1.0$). From $B$ to $C$, the copper hole content increases from $\approx 0.865$ to $\approx 0.883$ and correspondingly the oxygen hole content decreases from $\approx 0.135$ to $\approx 0.117$. The Cu moment $m_d^2$ follows the behavior of the copper hole content. 
Hence there are no qualitative differences between the path $A \rightarrow D$ and $B \rightarrow C$. As a result, the path $D \rightarrow C$ has a similar trend as the path $A \rightarrow B$. 

Note that the physical understanding of the paths $A \rightarrow D$ and $B \rightarrow C$ follows the interpretation discussed in Sec.~\ref{sec:zerodoping}. In other words, by keeping fixed the bare Cu-O energy distance $\tilde{\epsilon}_p -\epsilon_d$, an increase in $U_d$ suppresses the Cu double occupancy, and thus the Cu-O charge fluctuation. As a result, this lead to a redistribution of the electronic charge from Cu to O orbitals, and to an enhancement to the spin Cu local moment. 

Next we study the effect of doping on the charge redistribution among Cu and O orbitals. 
Fig.~\ref{fig:discussion}(c) shows the Cu hole content (red up triangles), O hole content (blue down triangles) and spin Cu local moment squared (green circles) evaluated at hole doping $\delta = 0.20$ (dashed lines) and electron doping $\delta=-0.20$ (dotted line) relative to the points $A, B, C, D$ in the Zaanen-Sawatzky-Allen diagram of Fig.~\ref{fig:discussion}(a).

Qualitatively, the charge redistribution for either hole or electron doping in Fig.~\ref{fig:discussion}(c) (dashed and dotted lines, respectively) follows that on the undoped system in Fig.~\ref{fig:discussion}(b). However, there are some quantitative differences between the hole and electron doping of the charge-transfer insulator. This is best seen along the path from $A$ to $B$ in Fig.~\ref{fig:discussion}(c): the copper hole content increases from $\approx 0.780$ to $\approx 0.986$ at $\delta=0.20$ and from $\approx 0.580$ to $\approx 0.676$ at $\delta = -0.20$. Thus, the increase is higher on hole doping. This reflects that in the charge-transfer regime, doped holes enter mainly the O orbitals, whereas doped electrons go mainly in the Cu orbital. This also implies that the slope of the curves on the hole doping side in Fig.~\ref{fig:choVSohc} increases upon decreasing $\tilde{\epsilon}_p -\epsilon_d$ (i.e. in going from the charge-transfer to the Mott-Hubbard regime), whereas it does not change substantially on the electron doping.

\subsection{Consequences for hole-doped cuprates}
\label{sec:discussion-exp}

Finally, we discuss how these results can indicate a possible direction toward understanding some trends in the cuprate phenomenology. 
Hole-doped cuprates are doped charge-transfer insulators, hence the relevant region of the Zaanen-Sawatzky-Allen diagram in Fig.~\ref{fig:discussion}(a) is described by the parameters $U_d > \Delta$ and $U_d > U_{d, {\rm MIT}}$. 
By combining the knowledge of the charge-transfer gap size and of the O hole content in the CuO$_2$ plane, we can place cuprate systems in the Zaanen-Sawatzky-Allen diagram and thus connect observed experimental trends to microscopic parameters of the Emery model. 
On one hand, the charge-transfer gap of the parent cuprates has been estimated by photoemission, STM and optical conductivity experiments to be in the range $\approx 1.5$ - $2$~eV~\cite{Damascelli:RMP2003}. 
On the other hand, the Cu (and thus the O) hole content in the parent $\delta=0$ state varies substantially among cuprates, ranging from $\approx 0.5$ to $\approx 0.85$ (corresponding to O hole content from $0.5$ to $0.15$), as found by NMR experiments~\cite{Jurkutat:PRB2014, Rybicki:NatComm2016, Jurkutat:PNAS2023}. 

We can understand how the values of the charge-transfer gap size and of the O hole content constrain the location of systems in the Zaanen-Sawatzky-Allen diagram by tracing in Fig.~\ref{fig:discussion}(a) two types of isolines: lines at constant charge gap size $\Delta _{\rm I} = 0.5, 1, 1.5$ (orange dot-dashed lines with down, left and right triangles) and lines at constant Cu hole content hole content $p_d = 0.7, 0.75, 0.8$, or equivalently O hole content $p_p = 0.3, 0.25, 0.2$ (cyan dashed lines with left, down and up triangles). 

As one possible example, systems with equal charge gap size but different Cu hole content lie at the intersection between these lines. In Fig.~\ref{fig:discussion}(a), the point labeled with 1 has a charge gap size $\Delta_{\rm I}=0.5$ and Cu hole content $0.8$ (O hole content $0.2$), whereas the point labeled with 2 has a charge gap size $\Delta_{\rm I}=0.5$ and Cu hole content $0.75$ (O hole content $0.25$). Therefore, systems with larger O hole content lie more deeply in the charge-transfer regime in the Zaanen-Sawatzky-Allen scheme, i.e. in regions of higher correlation strength $U_d$ and smaller bare charge-transfer energy $\Delta$. 

Let us try to obtain a physical intuition of this intriguing finding. On one hand, the hole per CuO$_2$ unit cell that localises in the charge-transfer insulator is shared among Cu and O orbitals. The degree of sharing of this hole is proportional to the Cu-O bond covalency. Hence, a large O hole content (i.e. a large sharing of this hole) corresponds to a large Cu-O bond covalency. 
On the other hand, the O content is inversely proportional to the charge-transfer gap size (see lines at constant $\tilde{\epsilon}_p -\epsilon_d$ in Fig.~\ref{fig:gap}(c)). 
Therefore, (i) the requirement of large O hole content, or equivalently large Cu-O bond covalency, sets the Emery model in the charge-transfer regime, at regions of small $\Delta$ in the Sawatzky-Allen scheme, and (ii) the requirement of small charge-transfer gap size sets the Emery model close to the metal-insulator boundary in the Sawatzky-Allen scheme. Taken together, these two requirements imply that cuprates with larger O hole content lie more deeply in the charge-transfer regime in the Zaanen-Sawatzky-Allen scheme, i.e. in regions of higher correlation strength $U_d$ and smaller bare charge-transfer energy $\Delta$. 

Note that our analysis complements well earlier cluster DMFT studies~\cite{go, StCyr:2025}. Specifically, Ref.~\cite{go}, which studies the closely related $d$-$p$ model with dynamical cluster approximation (DCA) at zero temperature, uses the experimental optical gap of undoped La$_{2-x}$Sr$_x$CuO$_4$ to constrain the bare Cu-O energy difference, adopting $U_d$ in a given range of values believed to be relevant for cuprates. In the context of our work, this corresponds to fixing the $y$-axis of the Zaanen-Sawatzky-Allen diagram within a given range, and to using the experimental charge gap size to constrain the values of $\Delta$ on the $x$-axis. 
Conversely, Ref.~\cite{StCyr:2025}, which studies the Emery model with CDMFT at zero temperature for a set of realistic cuprates parameters, uses a complementary approach: the experimental NMR results of O hole content~\cite{Jurkutat:PRB2014, Rybicki:NatComm2016} are used to constrain the values of $\Delta$. In the context of our work, we see that the experimental NMR O hole content constrains the value of $\Delta$ on the the $x$-axis of the Zaanen-Sawatzky-Allen diagram, but not those on the $y$-axis, since the cyan lines of constant O hole content are almost vertical in the $U_d-\Delta$ plane. 
Our work complements these two approaches. As in Ref.~\cite{StCyr:2025}, the value of O hole content provides one constraint in the Zaanen-Sawatzky-Allen diagram (cyan lines with triangles in Fig.~\ref{fig:discussion}(a)). As in Ref.~\cite{go}, the size of the charge gap provides another constraint on the Zaanen-Sawatzky-Allen diagram (orange lines with triangles in Fig.~\ref{fig:discussion}(a)). 

Our calculations thus show that the location of a system in the Zaanen-Sawatzky-Allen diagram can be constrained by the charge gap size and by the O hole content in the CuO$_2$ plane. Within this framework, clear microscopic trends emerge. As one possible example, we showed that systems with same charge gap size but different O hole content lie in different regions of the Zaanen-Sawatzky-Allen diagram, with systems at larger O hole content placed deeper in the charge transfer regime, i.e. in regions of higher $U_d$ and smaller $\Delta$. 

Note two main limitations of this approach: by focusing only on the $U_d$ and $\tilde{\epsilon}_p -\epsilon_d$ dependence of the charge redistribution in the CuO$_2$ plane, our calculations (i) ignore the differences between material-specific model parameters~\cite{Cedric:NatPhys2010, Weber2011, cedricApical, Cui:Science2022, BacqLabreuil:2025, Cui:NatComm2025}, and (ii) have not examined the role of other model parameters, including the interplay with the Cu-O hybridization $t_{pd}$, the next nearest-neighbor O-O hopping $t_{pp}'$ and the onsite Coulomb repulsion on the O orbitals $U_p$. In principle, $t_{pd}$ and $t'_{pp}$ can readily be tuned, although their impact on the Monte Carlo sign problem remain to be assessed. On the other hand, the inclusion of $U_p$ requires modifications of the CDMFT framework used in this work, since the oxygen orbitals cannot be integrated out. In particular, the segment formalism of CT-HYB cannot be used anymore in that case. Despite these limitations, we view the trends in our framework as a starting point for a better understanding of the physics of the Emery model. 

Furthermore, another possible interest of these findings derives from the proposal~\cite{Rybicki:NatComm2016, Jurkutat:PNAS2023} that the optimal superconducting transition temperature can be obtained by maximising the O hole content or by reducing the charge-transfer gap size (see also Ref.~\cite{Nicolas:PNAS2021}). Hence, our results would suggest that systems with high superconducting transition temperature should lie deeper in the charge-transfer regime and close to the metal-insulator boundary of the Zaanen-Sawatzky-Allen diagram, i.e. in regions of higher correlation strength $U_d$ and smaller bare charge-transfer energy $\Delta$. This is because systems in this regions satisfy the criteria of large O content and small charge-transfer gap. This current work thus provides a stepping stone for testing this hypothesis.  
For example, in the context of Fig.~\ref{fig:discussion}(a), this would imply that doping with holes a system at the point labeled by 2 in Fig.~\ref{fig:discussion}(a) would have a larger superconducting transition temperature than that obtained by doping with holes a system at the point labeled by 1.

\section{Summary and conclusions}
\label{sec:conclusion}

In summary, we have studied the electronic charge redistribution among Cu and O orbitals and the charge gap in the Emery model at finite temperature with CDMFT. The strategy of our work is to map out this charge redistribution in the Zaanen-Sawatzky-Allen scheme, i.e. as a function of the interaction strength $U_d$ and of the bare charge-transfer energy $\Delta$, all the way from the charge-transfer to the Mott-Hubbard regime, with the goal of identifying the microscopic mechanisms controlling its behavior. 
This charge redistribution allows us to make a link with NMR experiments in cuprates~\cite{Jurkutat:PRB2014, Rybicki:NatComm2016, Jurkutat:PNAS2023}. This charge redistribution is inaccessible in a one-band model.

At zero doping, we constructed the Zaanen-Sawatzky-Allen diagram (Fig.~\ref{fig:zsa}) and characterised the degree of sharing of the hole that localises in the CuO$_2$ unit cell (Fig.~\ref{fig:zero-doping}(a,b)), the local moment on the Cu orbital (Fig.~\ref{fig:zero-doping}(c)) and the size of the correlated gap (Fig.~\ref{fig:gap}). Next we quantified the charge redistribution (Fig.~\ref{fig:choVSohc}) and the Cu local moment (Fig.~\ref{fig:moment}) upon varying the electron and hole doping in the range $\delta \in [0, \pm 0.20]$. Then in Sec.~\ref{sec:discussion} we rationalised our results within the Zaanen-Sawatzky-Allen framework. 

The main finding of our work can be summarised as follows: the position relative to the metal-insulator boundary of the Zaanen-Sawatzky-Allen diagram governs the charge redistribution among Cu and O orbitals. 
Specifically, (i) at any doping level and at fixed bare Cu-O energy distance $\tilde{\epsilon}_p - \epsilon_d$, the Cu hole content increases with increasing $U_d$ at the expenses of the O hole content, all the way from the charge-transfer to the Mott-Hubbard regime (as summarised by paths $A \rightarrow D$ and $B \rightarrow C$ in Figs.~\ref{fig:discussion}(b),(c)). 
Physically, this is because a large $U_d$ costs large potential energy if electrons occupy the Cu orbital, hence favoring large hole content in the Cu orbital. 
(ii) At any doping level and for a fixed charge gap size of the parent state, as the system evolves from the charge-transfer to Mott-Hubbard regime, the Cu hole content increases at the expenses of the O hole content. Physically, this is due to the fact that the charge becomes more localised on the Cu orbitals (as summarised by paths $A \rightarrow B$ and $D \rightarrow C$ in Figs.~\ref{fig:discussion}(b),(c)).

These findings lead to the following conclusions. (i) $U_d$ and $\Delta$ are main drivers of the microscopic process of charge redistribution among Cu and O orbitals. 
(ii) The size of the correlated gap is inversely proportional to the O hole content, all the way from the charge-transfer to the Mott-Hubbard regime, supporting and extending earlier work~\cite{Nicolas:PNAS2021, Davis:PNAS2022, Jurkutat:PNAS2023}. 
(iii) The following two observables, the charge gap size and the partial occupancies, can be used to constrain the location of systems in the Zaanen-Sawatzky-Allen diagram. Previous work had suggested focusing either on charge transfer gap~\cite{go} or on O hole content~\cite{StCyr:2025} to constrain microscopic parameters. We have shown that we can use simultaneously these two strategies. Intriguingly, we found that systems with same charge gap size but different hole content lie in different regions of the Zaanen-Sawatzky-Allen diagram, with systems at larger O hole content lying in regions of higher $U_d$ and smaller $\Delta$. 
Hence, this work paves the way for verifying whether these regions optimises superconductivity, as proposed in the context of cuprates in Refs.~\cite{Rybicki:NatComm2016, Jurkutat:PNAS2023}.

\begin{acknowledgments}
We thank Patrick S\'emon for sharing his continuous-time quantum Monte Carlo code and Louis-Bernard St-Cyr and David S\'en\'echal for discussions. This work has been supported by the Canada First Research Excellence Fund. Simulations were performed on computers provided by the Canada Foundation for Innovation, the Minist\`ere de l'\'Education des Loisirs et du Sport (Qu\'ebec), Calcul Qu\'ebec, and Digital Research Alliance of Canada.
\end{acknowledgments}

\appendix 

\section{Average sign of the Monte Carlo simulations}
\label{appendix}

\begin{figure*}%[ht!]
\centering{
\includegraphics[width=1.0\linewidth]{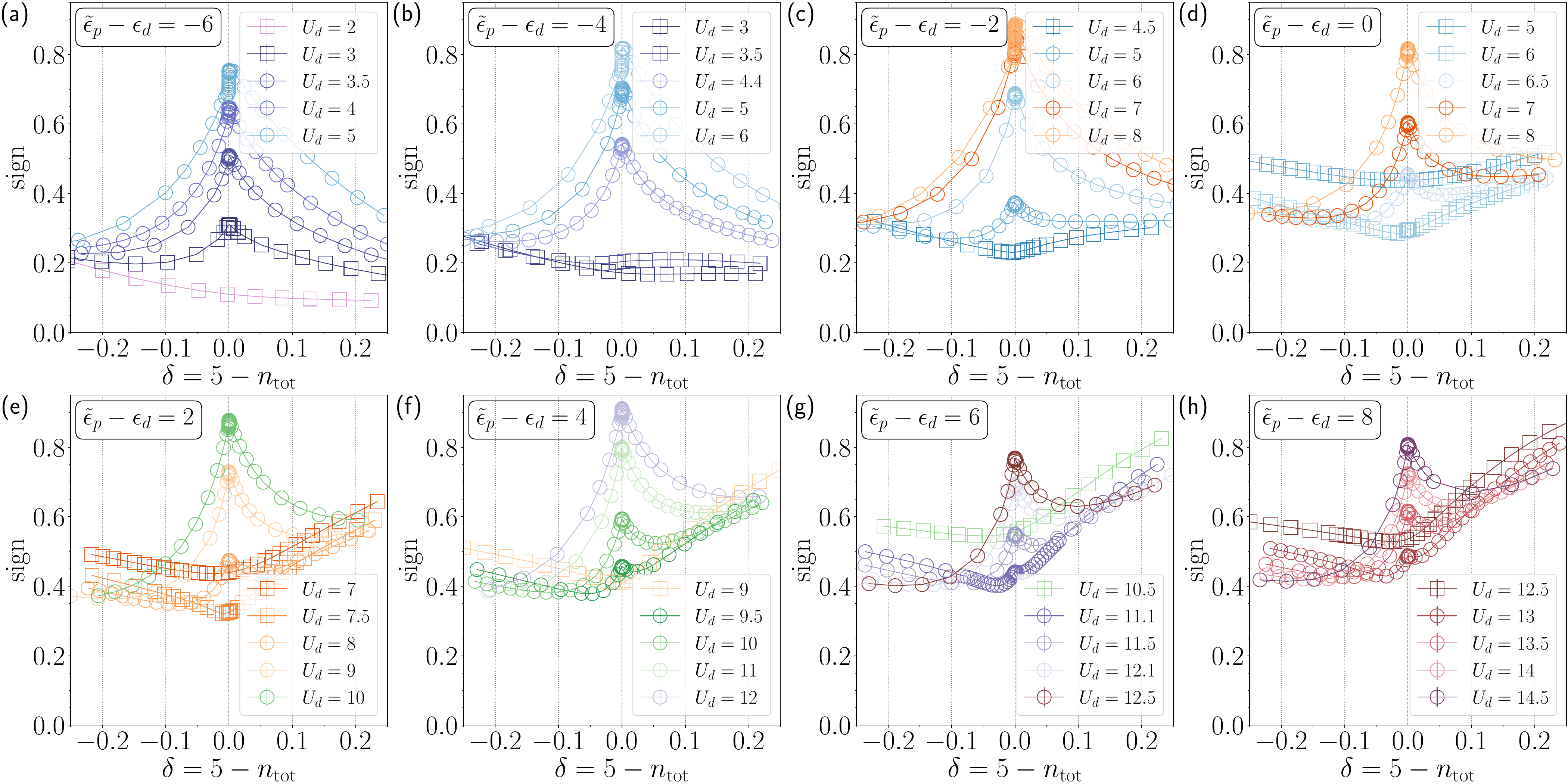}}
\caption{Monte Carlo sign vs $\delta$. Each panel shows data at a fixed value of the bare Cu-O energy distance $\tilde{\epsilon}_p -\epsilon_d$. Lines with circles (squares) indicate that the system at $\delta=0$ is a correlated insulator (metal). In each panel, the grey dashed line indicates $\delta=0$ and the grey dotted lines mark constant values of electron or hole doping. Other model parameters are as in Eq.~\ref{eq:SetParameters}.
}
\label{fig:sign}
\end{figure*}

Figure~\ref{fig:sign} shows the average sign of the continuous-time quantum Monte Carlo simulations as a function of doping $\delta$ and for different values of the bare Cu-O energy distance $\tilde{\epsilon}_p -\epsilon_d$, for the same values of the parameters in Fig.~\ref{fig:choVSohc}.

%

%\bibliography{emeryD}

\begin{thebibliography}{45}%
\makeatletter
\providecommand \@ifxundefined [1]{%
 \@ifx{#1\undefined}
}%
\providecommand \@ifnum [1]{%
 \ifnum #1\expandafter \@firstoftwo
 \else \expandafter \@secondoftwo
 \fi
}%
\providecommand \@ifx [1]{%
 \ifx #1\expandafter \@firstoftwo
 \else \expandafter \@secondoftwo
 \fi
}%
\providecommand \natexlab [1]{#1}%
\providecommand \enquote  [1]{``#1''}%
\providecommand \bibnamefont  [1]{#1}%
\providecommand \bibfnamefont [1]{#1}%
\providecommand \citenamefont [1]{#1}%
\providecommand \href@noop [0]{\@secondoftwo}%
\providecommand \href [0]{\begingroup \@sanitize@url \@href}%
\providecommand \@href[1]{\@@startlink{#1}\@@href}%
\providecommand \@@href[1]{\endgroup#1\@@endlink}%
\providecommand \@sanitize@url [0]{\catcode `\\12\catcode `\$12\catcode
  `\&12\catcode `\#12\catcode `\^12\catcode `\_12\catcode `\%12\relax}%
\providecommand \@@startlink[1]{}%
\providecommand \@@endlink[0]{}%
\providecommand \url  [0]{\begingroup\@sanitize@url \@url }%
\providecommand \@url [1]{\endgroup\@href {#1}{\urlprefix }}%
\providecommand \urlprefix  [0]{URL }%
\providecommand \Eprint [0]{\href }%
\providecommand \doibase [0]{https://doi.org/}%
\providecommand \selectlanguage [0]{\@gobble}%
\providecommand \bibinfo  [0]{\@secondoftwo}%
\providecommand \bibfield  [0]{\@secondoftwo}%
\providecommand \translation [1]{[#1]}%
\providecommand \BibitemOpen [0]{}%
\providecommand \bibitemStop [0]{}%
\providecommand \bibitemNoStop [0]{.\EOS\space}%
\providecommand \EOS [0]{\spacefactor3000\relax}%
\providecommand \BibitemShut  [1]{\csname bibitem#1\endcsname}%
\let\auto@bib@innerbib\@empty
%</preamble>
\bibitem [{\citenamefont {Keimer}\ \emph {et~al.}(2015)\citenamefont {Keimer},
  \citenamefont {Kivelson}, \citenamefont {Norman}, \citenamefont {Uchida},\
  and\ \citenamefont {Zaanen}}]{keimerRev}%
  \BibitemOpen
  \bibfield  {author} {\bibinfo {author} {\bibfnamefont {B.}~\bibnamefont
  {Keimer}}, \bibinfo {author} {\bibfnamefont {S.~A.}\ \bibnamefont
  {Kivelson}}, \bibinfo {author} {\bibfnamefont {M.~R.}\ \bibnamefont
  {Norman}}, \bibinfo {author} {\bibfnamefont {S.}~\bibnamefont {Uchida}},\
  and\ \bibinfo {author} {\bibfnamefont {J.}~\bibnamefont {Zaanen}},\
  }\bibfield  {title} {\bibinfo {title} {From quantum matter to
  high-temperature superconductivity in copper oxides},\ }\href
  {https://doi.org/10.1038/nature14165} {\bibfield  {journal} {\bibinfo
  {journal} {Nature}\ }\textbf {\bibinfo {volume} {518}},\ \bibinfo {pages}
  {179} (\bibinfo {year} {2015})}\BibitemShut {NoStop}%
\bibitem [{\citenamefont {Dagotto}(1994)}]{Dagotto:RMP1994}%
  \BibitemOpen
  \bibfield  {author} {\bibinfo {author} {\bibfnamefont {E.}~\bibnamefont
  {Dagotto}},\ }\bibfield  {title} {\bibinfo {title} {Correlated electrons in
  high-temperature superconductors},\ }\href
  {https://doi.org/10.1103/RevModPhys.66.763} {\bibfield  {journal} {\bibinfo
  {journal} {Rev. Mod. Phys.}\ }\textbf {\bibinfo {volume} {66}},\ \bibinfo
  {pages} {763} (\bibinfo {year} {1994})}\BibitemShut {NoStop}%
\bibitem [{\citenamefont {Imada}\ \emph {et~al.}(1998)\citenamefont {Imada},
  \citenamefont {Fujimori},\ and\ \citenamefont {Tokura}}]{ift}%
  \BibitemOpen
  \bibfield  {author} {\bibinfo {author} {\bibfnamefont {M.}~\bibnamefont
  {Imada}}, \bibinfo {author} {\bibfnamefont {A.}~\bibnamefont {Fujimori}},\
  and\ \bibinfo {author} {\bibfnamefont {Y.}~\bibnamefont {Tokura}},\
  }\bibfield  {title} {\bibinfo {title} {Metal-insulator transitions},\ }\href
  {https://doi.org/10.1103/RevModPhys.70.1039} {\bibfield  {journal} {\bibinfo
  {journal} {Rev. Mod. Phys.}\ }\textbf {\bibinfo {volume} {70}},\ \bibinfo
  {pages} {1039} (\bibinfo {year} {1998})}\BibitemShut {NoStop}%
\bibitem [{\citenamefont {{Rybicki}}\ \emph {et~al.}(2016)\citenamefont
  {{Rybicki}}, \citenamefont {{Jurkutat}}, \citenamefont {{Reichardt}},
  \citenamefont {{Kapusta}},\ and\ \citenamefont
  {{Haase}}}]{Rybicki:NatComm2016}%
  \BibitemOpen
  \bibfield  {author} {\bibinfo {author} {\bibfnamefont {D.}~\bibnamefont
  {{Rybicki}}}, \bibinfo {author} {\bibfnamefont {M.}~\bibnamefont
  {{Jurkutat}}}, \bibinfo {author} {\bibfnamefont {S.}~\bibnamefont
  {{Reichardt}}}, \bibinfo {author} {\bibfnamefont {C.}~\bibnamefont
  {{Kapusta}}},\ and\ \bibinfo {author} {\bibfnamefont {J.}~\bibnamefont
  {{Haase}}},\ }\bibfield  {title} {\bibinfo {title} {{Perspective on the phase
  diagram of cuprate high-temperature superconductors}},\ }\href
  {https://doi.org/10.1038/ncomms11413} {\bibfield  {journal} {\bibinfo
  {journal} {Nature Communications}\ }\textbf {\bibinfo {volume} {7}},\
  \bibinfo {eid} {11413} (\bibinfo {year} {2016})}\BibitemShut {NoStop}%
\bibitem [{\citenamefont {Jurkutat}\ \emph {et~al.}(2023)\citenamefont
  {Jurkutat}, \citenamefont {Kattinger}, \citenamefont {Tsankov}, \citenamefont
  {Reznicek}, \citenamefont {Erb},\ and\ \citenamefont
  {Haase}}]{Jurkutat:PNAS2023}%
  \BibitemOpen
  \bibfield  {author} {\bibinfo {author} {\bibfnamefont {M.}~\bibnamefont
  {Jurkutat}}, \bibinfo {author} {\bibfnamefont {C.}~\bibnamefont {Kattinger}},
  \bibinfo {author} {\bibfnamefont {S.}~\bibnamefont {Tsankov}}, \bibinfo
  {author} {\bibfnamefont {R.}~\bibnamefont {Reznicek}}, \bibinfo {author}
  {\bibfnamefont {A.}~\bibnamefont {Erb}},\ and\ \bibinfo {author}
  {\bibfnamefont {J.}~\bibnamefont {Haase}},\ }\bibfield  {title} {\bibinfo
  {title} {{How pressure enhances the critical temperature of superconductivity
  in YBa$_2$Cu$_3$O$_{6+y}$ }},\ }\href
  {https://doi.org/10.1073/pnas.2215458120} {\bibfield  {journal} {\bibinfo
  {journal} {Proceedings of the National Academy of Sciences}\ }\textbf
  {\bibinfo {volume} {120}},\ \bibinfo {pages} {e2215458120} (\bibinfo {year}
  {2023})}\BibitemShut {NoStop}%
\bibitem [{\citenamefont {Kowalski}\ \emph {et~al.}(2021)\citenamefont
  {Kowalski}, \citenamefont {Dash}, \citenamefont {Sémon}, \citenamefont
  {Sénéchal},\ and\ \citenamefont {Tremblay}}]{Nicolas:PNAS2021}%
  \BibitemOpen
  \bibfield  {author} {\bibinfo {author} {\bibfnamefont {N.}~\bibnamefont
  {Kowalski}}, \bibinfo {author} {\bibfnamefont {S.~S.}\ \bibnamefont {Dash}},
  \bibinfo {author} {\bibfnamefont {P.}~\bibnamefont {Sémon}}, \bibinfo
  {author} {\bibfnamefont {D.}~\bibnamefont {Sénéchal}},\ and\ \bibinfo
  {author} {\bibfnamefont {A.-M.}\ \bibnamefont {Tremblay}},\ }\bibfield
  {title} {\bibinfo {title} {Oxygen hole content, charge-transfer gap,
  covalency, and cuprate superconductivity},\ }\href
  {https://doi.org/10.1073/pnas.2106476118} {\bibfield  {journal} {\bibinfo
  {journal} {Proceedings of the National Academy of Sciences}\ }\textbf
  {\bibinfo {volume} {118}},\ \bibinfo {pages} {e2106476118} (\bibinfo {year}
  {2021})}\BibitemShut {NoStop}%
\bibitem [{\citenamefont {O’Mahony}\ \emph {et~al.}(2022)\citenamefont
  {O’Mahony}, \citenamefont {Ren}, \citenamefont {Chen}, \citenamefont
  {Chong}, \citenamefont {Liu}, \citenamefont {Eisaki}, \citenamefont {Uchida},
  \citenamefont {Hamidian},\ and\ \citenamefont {Davis}}]{Davis:PNAS2022}%
  \BibitemOpen
  \bibfield  {author} {\bibinfo {author} {\bibfnamefont {S.~M.}\ \bibnamefont
  {O’Mahony}}, \bibinfo {author} {\bibfnamefont {W.}~\bibnamefont {Ren}},
  \bibinfo {author} {\bibfnamefont {W.}~\bibnamefont {Chen}}, \bibinfo {author}
  {\bibfnamefont {Y.~X.}\ \bibnamefont {Chong}}, \bibinfo {author}
  {\bibfnamefont {X.}~\bibnamefont {Liu}}, \bibinfo {author} {\bibfnamefont
  {H.}~\bibnamefont {Eisaki}}, \bibinfo {author} {\bibfnamefont
  {S.}~\bibnamefont {Uchida}}, \bibinfo {author} {\bibfnamefont {M.~H.}\
  \bibnamefont {Hamidian}},\ and\ \bibinfo {author} {\bibfnamefont {J.~C.~S.}\
  \bibnamefont {Davis}},\ }\bibfield  {title} {\bibinfo {title} {On the
  electron pairing mechanism of copper-oxide high temperature
  superconductivity},\ }\href {https://doi.org/10.1073/pnas.2207449119}
  {\bibfield  {journal} {\bibinfo  {journal} {Proceedings of the National
  Academy of Sciences}\ }\textbf {\bibinfo {volume} {119}},\ \bibinfo {pages}
  {e2207449119} (\bibinfo {year} {2022})}\BibitemShut {NoStop}%
\bibitem [{\citenamefont {Emery}(1987)}]{Emery_1987}%
  \BibitemOpen
  \bibfield  {author} {\bibinfo {author} {\bibfnamefont {V.~J.}\ \bibnamefont
  {Emery}},\ }\bibfield  {title} {\bibinfo {title} {{ Theory of
  high-${\mathrm{T}}_{\mathrm{c}}$ superconductivity in oxides }},\ }\href
  {https://doi.org/10.1103/PhysRevLett.58.2794} {\bibfield  {journal} {\bibinfo
   {journal} {Phys. Rev. Lett.}\ }\textbf {\bibinfo {volume} {58}},\ \bibinfo
  {pages} {2794} (\bibinfo {year} {1987})}\BibitemShut {NoStop}%
\bibitem [{\citenamefont {Varma}\ \emph {et~al.}(1987)\citenamefont {Varma},
  \citenamefont {Schmitt-Rink},\ and\ \citenamefont {Abrahams}}]{Varma_1987}%
  \BibitemOpen
  \bibfield  {author} {\bibinfo {author} {\bibfnamefont {C.}~\bibnamefont
  {Varma}}, \bibinfo {author} {\bibfnamefont {S.}~\bibnamefont
  {Schmitt-Rink}},\ and\ \bibinfo {author} {\bibfnamefont {E.}~\bibnamefont
  {Abrahams}},\ }\bibfield  {title} {\bibinfo {title} {Charge transfer
  excitations and superconductivity in ionic metals},\ }\href
  {http://dx.doi.org/10.1016/0038-1098(87)90407-8} {\bibfield  {journal}
  {\bibinfo  {journal} {Solid State Communications}\ }\textbf {\bibinfo
  {volume} {62}},\ \bibinfo {pages} {681 } (\bibinfo {year}
  {1987})}\BibitemShut {NoStop}%
\bibitem [{\citenamefont {Zaanen}\ \emph {et~al.}(1985)\citenamefont {Zaanen},
  \citenamefont {Sawatzky},\ and\ \citenamefont {Allen}}]{zsa}%
  \BibitemOpen
  \bibfield  {author} {\bibinfo {author} {\bibfnamefont {J.}~\bibnamefont
  {Zaanen}}, \bibinfo {author} {\bibfnamefont {G.~A.}\ \bibnamefont
  {Sawatzky}},\ and\ \bibinfo {author} {\bibfnamefont {J.~W.}\ \bibnamefont
  {Allen}},\ }\bibfield  {title} {\bibinfo {title} {{ Band gaps and electronic
  structure of transition-metal compounds }},\ }\href
  {https://doi.org/10.1103/PhysRevLett.55.418} {\bibfield  {journal} {\bibinfo
  {journal} {Phys. Rev. Lett.}\ }\textbf {\bibinfo {volume} {55}},\ \bibinfo
  {pages} {418} (\bibinfo {year} {1985})}\BibitemShut {NoStop}%
\bibitem [{\citenamefont {Maier}\ \emph {et~al.}(2005)\citenamefont {Maier},
  \citenamefont {Jarrell}, \citenamefont {Pruschke},\ and\ \citenamefont
  {Hettler}}]{maier}%
  \BibitemOpen
  \bibfield  {author} {\bibinfo {author} {\bibfnamefont {T.}~\bibnamefont
  {Maier}}, \bibinfo {author} {\bibfnamefont {M.}~\bibnamefont {Jarrell}},
  \bibinfo {author} {\bibfnamefont {T.}~\bibnamefont {Pruschke}},\ and\
  \bibinfo {author} {\bibfnamefont {M.~H.}\ \bibnamefont {Hettler}},\
  }\bibfield  {title} {\bibinfo {title} {Quantum cluster theories},\ }\href
  {https://doi.org/10.1103/RevModPhys.77.1027} {\bibfield  {journal} {\bibinfo
  {journal} {Rev. Mod. Phys.}\ }\textbf {\bibinfo {volume} {77}},\ \bibinfo
  {pages} {1027} (\bibinfo {year} {2005})}\BibitemShut {NoStop}%
\bibitem [{\citenamefont {Kotliar}\ \emph {et~al.}(2006)\citenamefont
  {Kotliar}, \citenamefont {Savrasov}, \citenamefont {Haule}, \citenamefont
  {Oudovenko}, \citenamefont {Parcollet},\ and\ \citenamefont
  {Marianetti}}]{kotliarRMP}%
  \BibitemOpen
  \bibfield  {author} {\bibinfo {author} {\bibfnamefont {G.}~\bibnamefont
  {Kotliar}}, \bibinfo {author} {\bibfnamefont {S.~Y.}\ \bibnamefont
  {Savrasov}}, \bibinfo {author} {\bibfnamefont {K.}~\bibnamefont {Haule}},
  \bibinfo {author} {\bibfnamefont {V.~S.}\ \bibnamefont {Oudovenko}}, \bibinfo
  {author} {\bibfnamefont {O.}~\bibnamefont {Parcollet}},\ and\ \bibinfo
  {author} {\bibfnamefont {C.~A.}\ \bibnamefont {Marianetti}},\ }\bibfield
  {title} {\bibinfo {title} {{Electronic structure calculations with dynamical
  mean-field theory}},\ }\href {https://doi.org/10.1103/RevModPhys.78.865}
  {\bibfield  {journal} {\bibinfo  {journal} {Rev. Mod. Phys.}\ }\textbf
  {\bibinfo {volume} {78}},\ \bibinfo {eid} {865} (\bibinfo {year}
  {2006})}\BibitemShut {NoStop}%
\bibitem [{\citenamefont {Tremblay}\ \emph {et~al.}(2006)\citenamefont
  {Tremblay}, \citenamefont {Kyung},\ and\ \citenamefont
  {S\'{e}n\'{e}chal}}]{tremblayR}%
  \BibitemOpen
  \bibfield  {author} {\bibinfo {author} {\bibfnamefont {A.-M.~S.}\
  \bibnamefont {Tremblay}}, \bibinfo {author} {\bibfnamefont {B.}~\bibnamefont
  {Kyung}},\ and\ \bibinfo {author} {\bibfnamefont {D.}~\bibnamefont
  {S\'{e}n\'{e}chal}},\ }\bibfield  {title} {\bibinfo {title} {{Pseudogap and
  high-temperature superconductivity from weak to strong coupling. Towards a
  quantitative theory}},\ }\href {https://doi.org/10.1063/1.2199446} {\bibfield
   {journal} {\bibinfo  {journal} {Low Temp. Phys.}\ }\textbf {\bibinfo
  {volume} {32}},\ \bibinfo {pages} {424} (\bibinfo {year} {2006})}\BibitemShut
  {NoStop}%
\bibitem [{\citenamefont {Georges}\ \emph {et~al.}(1996)\citenamefont
  {Georges}, \citenamefont {Kotliar}, \citenamefont {Krauth},\ and\
  \citenamefont {Rozenberg}}]{rmp}%
  \BibitemOpen
  \bibfield  {author} {\bibinfo {author} {\bibfnamefont {A.}~\bibnamefont
  {Georges}}, \bibinfo {author} {\bibfnamefont {G.}~\bibnamefont {Kotliar}},
  \bibinfo {author} {\bibfnamefont {W.}~\bibnamefont {Krauth}},\ and\ \bibinfo
  {author} {\bibfnamefont {M.~J.}\ \bibnamefont {Rozenberg}},\ }\bibfield
  {title} {\bibinfo {title} {{Dynamical mean-field theory of strongly
  correlated fermion systems and the limit of infinite dimensions}},\ }\href
  {https://doi.org/10.1103/RevModPhys.68.13} {\bibfield  {journal} {\bibinfo
  {journal} {Rev. Mod. Phys.}\ }\textbf {\bibinfo {volume} {68}},\ \bibinfo
  {pages} {13} (\bibinfo {year} {1996})}\BibitemShut {NoStop}%
\bibitem [{\citenamefont {Andersen}\ \emph {et~al.}(1995)\citenamefont
  {Andersen}, \citenamefont {Liechtenstein}, \citenamefont {Jepsen},\ and\
  \citenamefont {Paulsen}}]{AndersenLDA}%
  \BibitemOpen
  \bibfield  {author} {\bibinfo {author} {\bibfnamefont {O.}~\bibnamefont
  {Andersen}}, \bibinfo {author} {\bibfnamefont {A.}~\bibnamefont
  {Liechtenstein}}, \bibinfo {author} {\bibfnamefont {O.}~\bibnamefont
  {Jepsen}},\ and\ \bibinfo {author} {\bibfnamefont {F.}~\bibnamefont
  {Paulsen}},\ }\bibfield  {title} {\bibinfo {title} {{LDA energy bands, low
  energy hamiltonians, t',t'', $t_{\perp}(k)$ and $J_{\perp}$}},\ }\href
  {https://doi.org/http://dx.doi.org/10.1016/0022-3697(95)00269-3} {\bibfield
  {journal} {\bibinfo  {journal} {J. Phys, Chem. Solids}\ }\textbf {\bibinfo
  {volume} {56}},\ \bibinfo {pages} {1573} (\bibinfo {year}
  {1995})}\BibitemShut {NoStop}%
\bibitem [{\citenamefont {Gull}\ \emph {et~al.}(2011)\citenamefont {Gull},
  \citenamefont {Millis}, \citenamefont {Lichtenstein}, \citenamefont
  {Rubtsov}, \citenamefont {Troyer},\ and\ \citenamefont {Werner}}]{millisRMP}%
  \BibitemOpen
  \bibfield  {author} {\bibinfo {author} {\bibfnamefont {E.}~\bibnamefont
  {Gull}}, \bibinfo {author} {\bibfnamefont {A.~J.}\ \bibnamefont {Millis}},
  \bibinfo {author} {\bibfnamefont {A.~I.}\ \bibnamefont {Lichtenstein}},
  \bibinfo {author} {\bibfnamefont {A.~N.}\ \bibnamefont {Rubtsov}}, \bibinfo
  {author} {\bibfnamefont {M.}~\bibnamefont {Troyer}},\ and\ \bibinfo {author}
  {\bibfnamefont {P.}~\bibnamefont {Werner}},\ }\bibfield  {title} {\bibinfo
  {title} {{Continuous-time Monte~Carlo methods for quantum impurity models}},\
  }\href {https://doi.org/10.1103/RevModPhys.83.349} {\bibfield  {journal}
  {\bibinfo  {journal} {Rev. Mod. Phys.}\ }\textbf {\bibinfo {volume} {83}},\
  \bibinfo {pages} {349} (\bibinfo {year} {2011})}\BibitemShut {NoStop}%
\bibitem [{\citenamefont {Werner}\ \emph {et~al.}(2006)\citenamefont {Werner},
  \citenamefont {Comanac}, \citenamefont {de~Medici}, \citenamefont {Troyer},\
  and\ \citenamefont {Millis}}]{Werner:2006}%
  \BibitemOpen
  \bibfield  {author} {\bibinfo {author} {\bibfnamefont {P.}~\bibnamefont
  {Werner}}, \bibinfo {author} {\bibfnamefont {A.}~\bibnamefont {Comanac}},
  \bibinfo {author} {\bibfnamefont {L.}~\bibnamefont {de~Medici}}, \bibinfo
  {author} {\bibfnamefont {M.}~\bibnamefont {Troyer}},\ and\ \bibinfo {author}
  {\bibfnamefont {A.~J.}\ \bibnamefont {Millis}},\ }\bibfield  {title}
  {\bibinfo {title} {Continuous-time solver for quantum impurity models},\
  }\href {https://doi.org/10.1103/PhysRevLett.97.076405} {\bibfield  {journal}
  {\bibinfo  {journal} {Phys. Rev. Lett.}\ }\textbf {\bibinfo {volume} {97}},\
  \bibinfo {pages} {076405} (\bibinfo {year} {2006})}\BibitemShut {NoStop}%
\bibitem [{\citenamefont {Haule}(2007)}]{hauleCTQMC}%
  \BibitemOpen
  \bibfield  {author} {\bibinfo {author} {\bibfnamefont {K.}~\bibnamefont
  {Haule}},\ }\bibfield  {title} {\bibinfo {title} {{Quantum Monte Carlo
  impurity solver for cluster dynamical mean-field theory and electronic
  structure calculations with adjustable cluster base}},\ }\href
  {https://doi.org/10.1103/PhysRevB.75.155113} {\bibfield  {journal} {\bibinfo
  {journal} {Phys. Rev. B}\ }\textbf {\bibinfo {volume} {75}},\ \bibinfo {eid}
  {155113} (\bibinfo {year} {2007})}\BibitemShut {NoStop}%
\bibitem [{\citenamefont {S\'emon}\ \emph {et~al.}(2014)\citenamefont
  {S\'emon}, \citenamefont {Yee}, \citenamefont {Haule},\ and\ \citenamefont
  {Tremblay}}]{patrickSkipList}%
  \BibitemOpen
  \bibfield  {author} {\bibinfo {author} {\bibfnamefont {P.}~\bibnamefont
  {S\'emon}}, \bibinfo {author} {\bibfnamefont {C.-H.}\ \bibnamefont {Yee}},
  \bibinfo {author} {\bibfnamefont {K.}~\bibnamefont {Haule}},\ and\ \bibinfo
  {author} {\bibfnamefont {A.-M.~S.}\ \bibnamefont {Tremblay}},\ }\bibfield
  {title} {\bibinfo {title} {{Lazy skip-lists: An algorithm for fast
  hybridization-expansion quantum Monte Carlo}},\ }\href
  {https://doi.org/10.1103/PhysRevB.90.075149} {\bibfield  {journal} {\bibinfo
  {journal} {Phys. Rev. B}\ }\textbf {\bibinfo {volume} {90}},\ \bibinfo
  {pages} {075149} (\bibinfo {year} {2014})}\BibitemShut {NoStop}%
\bibitem [{\citenamefont {Fratino}\ \emph {et~al.}(2016)\citenamefont
  {Fratino}, \citenamefont {S\'emon}, \citenamefont {Sordi},\ and\
  \citenamefont {Tremblay}}]{Lorenzo3band}%
  \BibitemOpen
  \bibfield  {author} {\bibinfo {author} {\bibfnamefont {L.}~\bibnamefont
  {Fratino}}, \bibinfo {author} {\bibfnamefont {P.}~\bibnamefont {S\'emon}},
  \bibinfo {author} {\bibfnamefont {G.}~\bibnamefont {Sordi}},\ and\ \bibinfo
  {author} {\bibfnamefont {A.-M.~S.}\ \bibnamefont {Tremblay}},\ }\bibfield
  {title} {\bibinfo {title} {Pseudogap and superconductivity in two-dimensional
  doped charge-transfer insulators},\ }\href
  {https://doi.org/10.1103/PhysRevB.93.245147} {\bibfield  {journal} {\bibinfo
  {journal} {Phys. Rev. B}\ }\textbf {\bibinfo {volume} {93}},\ \bibinfo
  {pages} {245147} (\bibinfo {year} {2016})}\BibitemShut {NoStop}%
\bibitem [{\citenamefont {Kowalski}(2021)}]{Nicolas:Master}%
  \BibitemOpen
  \bibfield  {author} {\bibinfo {author} {\bibfnamefont {N.}~\bibnamefont
  {Kowalski}},\ }\emph {\bibinfo {title} {{Dopage, temperature critique et
  \'etude du mod\`ele de Hubbard \`a trois bandes}}},\ \href@noop {} {Master's
  thesis},\ \bibinfo  {school} {Universit\'e de Sherbrooke}, \bibinfo {address}
  {Sherbrooke, QC, Canada} (\bibinfo {year} {2021})\BibitemShut {NoStop}%
\bibitem [{\citenamefont {Sordi}\ \emph {et~al.}(2025)\citenamefont {Sordi},
  \citenamefont {Reaney}, \citenamefont {Kowalski}, \citenamefont {S\'emon},\
  and\ \citenamefont {Tremblay}}]{GiovanniPRB2025}%
  \BibitemOpen
  \bibfield  {author} {\bibinfo {author} {\bibfnamefont {G.}~\bibnamefont
  {Sordi}}, \bibinfo {author} {\bibfnamefont {G.~L.}\ \bibnamefont {Reaney}},
  \bibinfo {author} {\bibfnamefont {N.}~\bibnamefont {Kowalski}}, \bibinfo
  {author} {\bibfnamefont {P.}~\bibnamefont {S\'emon}},\ and\ \bibinfo {author}
  {\bibfnamefont {A.-M.~S.}\ \bibnamefont {Tremblay}},\ }\bibfield  {title}
  {\bibinfo {title} {{Ambipolar doping of a charge-transfer insulator in the
  Emery model}},\ }\href {https://doi.org/10.1103/PhysRevB.111.045117}
  {\bibfield  {journal} {\bibinfo  {journal} {Phys. Rev. B}\ }\textbf {\bibinfo
  {volume} {111}},\ \bibinfo {pages} {045117} (\bibinfo {year}
  {2025})}\BibitemShut {NoStop}%
\bibitem [{\citenamefont {Park}\ \emph {et~al.}(2008)\citenamefont {Park},
  \citenamefont {Haule},\ and\ \citenamefont {Kotliar}}]{phk}%
  \BibitemOpen
  \bibfield  {author} {\bibinfo {author} {\bibfnamefont {H.}~\bibnamefont
  {Park}}, \bibinfo {author} {\bibfnamefont {K.}~\bibnamefont {Haule}},\ and\
  \bibinfo {author} {\bibfnamefont {G.}~\bibnamefont {Kotliar}},\ }\bibfield
  {title} {\bibinfo {title} {{Cluster Dynamical Mean Field Theory of the Mott
  Transition}},\ }\href {https://doi.org/10.1103/PhysRevLett.101.186403}
  {\bibfield  {journal} {\bibinfo  {journal} {Phys. Rev. Lett.}\ }\textbf
  {\bibinfo {volume} {101}},\ \bibinfo {eid} {186403} (\bibinfo {year}
  {2008})}\BibitemShut {NoStop}%
\bibitem [{\citenamefont {Go}\ and\ \citenamefont {Millis}(2015)}]{go}%
  \BibitemOpen
  \bibfield  {author} {\bibinfo {author} {\bibfnamefont {A.}~\bibnamefont
  {Go}}\ and\ \bibinfo {author} {\bibfnamefont {A.~J.}\ \bibnamefont
  {Millis}},\ }\bibfield  {title} {\bibinfo {title} {Spatial correlations and
  the insulating phase of the high-${T}_{c}$ cuprates: Insights from a
  configuration-interaction-based solver for dynamical mean field theory},\
  }\href {https://doi.org/10.1103/PhysRevLett.114.016402} {\bibfield  {journal}
  {\bibinfo  {journal} {Phys. Rev. Lett.}\ }\textbf {\bibinfo {volume} {114}},\
  \bibinfo {pages} {016402} (\bibinfo {year} {2015})}\BibitemShut {NoStop}%
\bibitem [{\citenamefont {Bergeron}\ and\ \citenamefont
  {Tremblay}(2016)}]{DominicMEM}%
  \BibitemOpen
  \bibfield  {author} {\bibinfo {author} {\bibfnamefont {D.}~\bibnamefont
  {Bergeron}}\ and\ \bibinfo {author} {\bibfnamefont {A.-M.~S.}\ \bibnamefont
  {Tremblay}},\ }\bibfield  {title} {\bibinfo {title} {Algorithms for optimized
  maximum entropy and diagnostic tools for analytic continuation},\ }\href
  {https://doi.org/10.1103/PhysRevE.94.023303} {\bibfield  {journal} {\bibinfo
  {journal} {Phys. Rev. E}\ }\textbf {\bibinfo {volume} {94}},\ \bibinfo
  {pages} {023303} (\bibinfo {year} {2016})}\BibitemShut {NoStop}%
\bibitem [{\citenamefont {Scalettar}\ \emph {et~al.}(1991)\citenamefont
  {Scalettar}, \citenamefont {Scalapino}, \citenamefont {Sugar},\ and\
  \citenamefont {White}}]{Scalettar:PRB1991}%
  \BibitemOpen
  \bibfield  {author} {\bibinfo {author} {\bibfnamefont {R.~T.}\ \bibnamefont
  {Scalettar}}, \bibinfo {author} {\bibfnamefont {D.~J.}\ \bibnamefont
  {Scalapino}}, \bibinfo {author} {\bibfnamefont {R.~L.}\ \bibnamefont
  {Sugar}},\ and\ \bibinfo {author} {\bibfnamefont {S.~R.}\ \bibnamefont
  {White}},\ }\bibfield  {title} {\bibinfo {title} {{Antiferromagnetic,
  charge-transfer, and pairing correlations in the three-band Hubbard model}},\
  }\href {https://doi.org/10.1103/PhysRevB.44.770} {\bibfield  {journal}
  {\bibinfo  {journal} {Phys. Rev. B}\ }\textbf {\bibinfo {volume} {44}},\
  \bibinfo {pages} {770} (\bibinfo {year} {1991})}\BibitemShut {NoStop}%
\bibitem [{\citenamefont {Arrigoni}\ \emph {et~al.}(2009)\citenamefont
  {Arrigoni}, \citenamefont {Aichhorn}, \citenamefont {Daghofer},\ and\
  \citenamefont {Hanke}}]{ArrigoniCuO2}%
  \BibitemOpen
  \bibfield  {author} {\bibinfo {author} {\bibfnamefont {E.}~\bibnamefont
  {Arrigoni}}, \bibinfo {author} {\bibfnamefont {M.}~\bibnamefont {Aichhorn}},
  \bibinfo {author} {\bibfnamefont {M.}~\bibnamefont {Daghofer}},\ and\
  \bibinfo {author} {\bibfnamefont {W.}~\bibnamefont {Hanke}},\ }\bibfield
  {title} {\bibinfo {title} {{Phase diagram and single-particle spectrum of
  $\textrm{CuO}_2$ high- $\textrm{T}_c$ layers: variational cluster approach to
  the three-band Hubbard model}},\ }\href
  {http://stacks.iop.org/1367-2630/11/i=5/a=055066} {\bibfield  {journal}
  {\bibinfo  {journal} {New Journal of Physics}\ }\textbf {\bibinfo {volume}
  {11}},\ \bibinfo {pages} {055066} (\bibinfo {year} {2009})}\BibitemShut
  {NoStop}%
\bibitem [{\citenamefont {Cui}\ \emph {et~al.}(2020)\citenamefont {Cui},
  \citenamefont {Sun}, \citenamefont {Ray}, \citenamefont {Zheng},
  \citenamefont {Sun},\ and\ \citenamefont {Chan}}]{Cui:PRR2020}%
  \BibitemOpen
  \bibfield  {author} {\bibinfo {author} {\bibfnamefont {Z.-H.}\ \bibnamefont
  {Cui}}, \bibinfo {author} {\bibfnamefont {C.}~\bibnamefont {Sun}}, \bibinfo
  {author} {\bibfnamefont {U.}~\bibnamefont {Ray}}, \bibinfo {author}
  {\bibfnamefont {B.-X.}\ \bibnamefont {Zheng}}, \bibinfo {author}
  {\bibfnamefont {Q.}~\bibnamefont {Sun}},\ and\ \bibinfo {author}
  {\bibfnamefont {G.~K.-L.}\ \bibnamefont {Chan}},\ }\bibfield  {title}
  {\bibinfo {title} {{Ground-state phase diagram of the three-band Hubbard
  model from density matrix embedding theory}},\ }\href
  {https://doi.org/10.1103/PhysRevResearch.2.043259} {\bibfield  {journal}
  {\bibinfo  {journal} {Phys. Rev. Res.}\ }\textbf {\bibinfo {volume} {2}},\
  \bibinfo {pages} {043259} (\bibinfo {year} {2020})}\BibitemShut {NoStop}%
\bibitem [{\citenamefont {Anderson}(1950)}]{AndersonSE}%
  \BibitemOpen
  \bibfield  {author} {\bibinfo {author} {\bibfnamefont {P.~W.}\ \bibnamefont
  {Anderson}},\ }\bibfield  {title} {\bibinfo {title} {{Antiferromagnetism.
  Theory of Superexchange Interaction}},\ }\href
  {https://doi.org/10.1103/PhysRev.79.350} {\bibfield  {journal} {\bibinfo
  {journal} {Phys. Rev.}\ }\textbf {\bibinfo {volume} {79}},\ \bibinfo {pages}
  {350} (\bibinfo {year} {1950})}\BibitemShut {NoStop}%
\bibitem [{\citenamefont {Anderson}(1987)}]{Anderson:1987}%
  \BibitemOpen
  \bibfield  {author} {\bibinfo {author} {\bibfnamefont {P.~W.}\ \bibnamefont
  {Anderson}},\ }\bibfield  {title} {\bibinfo {title} {{The resonating valence
  bond state in La$_2$CuO$_4$ and superconductivity}},\ }\href
  {https://doi.org/10.1126/science.235.4793.1196} {\bibfield  {journal}
  {\bibinfo  {journal} {Science}\ }\textbf {\bibinfo {volume} {235}},\ \bibinfo
  {pages} {1196} (\bibinfo {year} {1987})}\BibitemShut {NoStop}%
\bibitem [{\citenamefont {Mermin}\ and\ \citenamefont
  {Wagner}(1966)}]{MWtheorem}%
  \BibitemOpen
  \bibfield  {author} {\bibinfo {author} {\bibfnamefont {N.~D.}\ \bibnamefont
  {Mermin}}\ and\ \bibinfo {author} {\bibfnamefont {H.}~\bibnamefont
  {Wagner}},\ }\bibfield  {title} {\bibinfo {title} {{Absence of Ferromagnetism
  or Antiferromagnetism in One- or Two-Dimensional Isotropic Heisenberg
  Models}},\ }\href {https://doi.org/10.1103/PhysRevLett.17.1133} {\bibfield
  {journal} {\bibinfo  {journal} {Phys. Rev. Lett.}\ }\textbf {\bibinfo
  {volume} {17}},\ \bibinfo {pages} {1133} (\bibinfo {year}
  {1966})}\BibitemShut {NoStop}%
\bibitem [{\citenamefont {Kyung}\ \emph {et~al.}(2006)\citenamefont {Kyung},
  \citenamefont {Kancharla}, \citenamefont {S\'{e}n\'{e}chal}, \citenamefont
  {Tremblay}, \citenamefont {Civelli},\ and\ \citenamefont {Kotliar}}]{kyung}%
  \BibitemOpen
  \bibfield  {author} {\bibinfo {author} {\bibfnamefont {B.}~\bibnamefont
  {Kyung}}, \bibinfo {author} {\bibfnamefont {S.~S.}\ \bibnamefont
  {Kancharla}}, \bibinfo {author} {\bibfnamefont {D.}~\bibnamefont
  {S\'{e}n\'{e}chal}}, \bibinfo {author} {\bibfnamefont {A.-M.~S.}\
  \bibnamefont {Tremblay}}, \bibinfo {author} {\bibfnamefont {M.}~\bibnamefont
  {Civelli}},\ and\ \bibinfo {author} {\bibfnamefont {G.}~\bibnamefont
  {Kotliar}},\ }\bibfield  {title} {\bibinfo {title} {{Pseudogap induced by
  short-range spin correlations in a doped Mott insulator}},\ }\href
  {https://doi.org/10.1103/PhysRevB.73.165114} {\bibfield  {journal} {\bibinfo
  {journal} {Phys. Rev. B}\ }\textbf {\bibinfo {volume} {73}},\ \bibinfo {eid}
  {165114} (\bibinfo {year} {2006})}\BibitemShut {NoStop}%
\bibitem [{\citenamefont {Jurkutat}\ \emph {et~al.}(2014)\citenamefont
  {Jurkutat}, \citenamefont {Rybicki}, \citenamefont {Sushkov}, \citenamefont
  {Williams}, \citenamefont {Erb},\ and\ \citenamefont
  {Haase}}]{Jurkutat:PRB2014}%
  \BibitemOpen
  \bibfield  {author} {\bibinfo {author} {\bibfnamefont {M.}~\bibnamefont
  {Jurkutat}}, \bibinfo {author} {\bibfnamefont {D.}~\bibnamefont {Rybicki}},
  \bibinfo {author} {\bibfnamefont {O.~P.}\ \bibnamefont {Sushkov}}, \bibinfo
  {author} {\bibfnamefont {G.~V.~M.}\ \bibnamefont {Williams}}, \bibinfo
  {author} {\bibfnamefont {A.}~\bibnamefont {Erb}},\ and\ \bibinfo {author}
  {\bibfnamefont {J.}~\bibnamefont {Haase}},\ }\bibfield  {title} {\bibinfo
  {title} {Distribution of electrons and holes in cuprate superconductors as
  determined from ${}^{17}\mathrm{O}$ and $^{63}\mathrm{Cu}$ nuclear magnetic
  resonance},\ }\href {https://doi.org/10.1103/PhysRevB.90.140504} {\bibfield
  {journal} {\bibinfo  {journal} {Phys. Rev. B}\ }\textbf {\bibinfo {volume}
  {90}},\ \bibinfo {pages} {140504} (\bibinfo {year} {2014})}\BibitemShut
  {NoStop}%
\bibitem [{\citenamefont {White}\ and\ \citenamefont
  {Scalapino}(2015)}]{White:PRB2015}%
  \BibitemOpen
  \bibfield  {author} {\bibinfo {author} {\bibfnamefont {S.~R.}\ \bibnamefont
  {White}}\ and\ \bibinfo {author} {\bibfnamefont {D.~J.}\ \bibnamefont
  {Scalapino}},\ }\bibfield  {title} {\bibinfo {title} {{Doping asymmetry and
  striping in a three-orbital ${\mathrm{CuO}}_{2}$ Hubbard model}},\ }\href
  {https://doi.org/10.1103/PhysRevB.92.205112} {\bibfield  {journal} {\bibinfo
  {journal} {Phys. Rev. B}\ }\textbf {\bibinfo {volume} {92}},\ \bibinfo
  {pages} {205112} (\bibinfo {year} {2015})}\BibitemShut {NoStop}%
\bibitem [{\citenamefont {Kung}\ \emph {et~al.}(2016)\citenamefont {Kung},
  \citenamefont {Chen}, \citenamefont {Wang}, \citenamefont {Huang},
  \citenamefont {Nowadnick}, \citenamefont {Moritz}, \citenamefont {Scalettar},
  \citenamefont {Johnston},\ and\ \citenamefont {Devereaux}}]{Kung:PRB2016}%
  \BibitemOpen
  \bibfield  {author} {\bibinfo {author} {\bibfnamefont {Y.~F.}\ \bibnamefont
  {Kung}}, \bibinfo {author} {\bibfnamefont {C.-C.}\ \bibnamefont {Chen}},
  \bibinfo {author} {\bibfnamefont {Y.}~\bibnamefont {Wang}}, \bibinfo {author}
  {\bibfnamefont {E.~W.}\ \bibnamefont {Huang}}, \bibinfo {author}
  {\bibfnamefont {E.~A.}\ \bibnamefont {Nowadnick}}, \bibinfo {author}
  {\bibfnamefont {B.}~\bibnamefont {Moritz}}, \bibinfo {author} {\bibfnamefont
  {R.~T.}\ \bibnamefont {Scalettar}}, \bibinfo {author} {\bibfnamefont
  {S.}~\bibnamefont {Johnston}},\ and\ \bibinfo {author} {\bibfnamefont
  {T.~P.}\ \bibnamefont {Devereaux}},\ }\bibfield  {title} {\bibinfo {title}
  {{Characterizing the three-orbital Hubbard model with determinant quantum
  Monte Carlo}},\ }\href {https://doi.org/10.1103/PhysRevB.93.155166}
  {\bibfield  {journal} {\bibinfo  {journal} {Phys. Rev. B}\ }\textbf {\bibinfo
  {volume} {93}},\ \bibinfo {pages} {155166} (\bibinfo {year}
  {2016})}\BibitemShut {NoStop}%
\bibitem [{\citenamefont {Ponsioen}\ \emph {et~al.}(2023)\citenamefont
  {Ponsioen}, \citenamefont {Chung},\ and\ \citenamefont
  {Corboz}}]{Ponsioen:PRB2023}%
  \BibitemOpen
  \bibfield  {author} {\bibinfo {author} {\bibfnamefont {B.}~\bibnamefont
  {Ponsioen}}, \bibinfo {author} {\bibfnamefont {S.~S.}\ \bibnamefont
  {Chung}},\ and\ \bibinfo {author} {\bibfnamefont {P.}~\bibnamefont
  {Corboz}},\ }\bibfield  {title} {\bibinfo {title} {{Superconducting stripes
  in the hole-doped three-band Hubbard model}},\ }\href
  {https://doi.org/10.1103/PhysRevB.108.205154} {\bibfield  {journal} {\bibinfo
   {journal} {Phys. Rev. B}\ }\textbf {\bibinfo {volume} {108}},\ \bibinfo
  {pages} {205154} (\bibinfo {year} {2023})}\BibitemShut {NoStop}%
\bibitem [{\citenamefont {Mai}\ \emph {et~al.}(2024)\citenamefont {Mai},
  \citenamefont {Cohen-Stead}, \citenamefont {Maier},\ and\ \citenamefont
  {Johnston}}]{Mai:PNAS2024}%
  \BibitemOpen
  \bibfield  {author} {\bibinfo {author} {\bibfnamefont {P.}~\bibnamefont
  {Mai}}, \bibinfo {author} {\bibfnamefont {B.}~\bibnamefont {Cohen-Stead}},
  \bibinfo {author} {\bibfnamefont {T.~A.}\ \bibnamefont {Maier}},\ and\
  \bibinfo {author} {\bibfnamefont {S.}~\bibnamefont {Johnston}},\ }\bibfield
  {title} {\bibinfo {title} {{Fluctuating charge-density-wave correlations in
  the three-band Hubbard model}},\ }\href
  {https://doi.org/10.1073/pnas.2408717121} {\bibfield  {journal} {\bibinfo
  {journal} {Proceedings of the National Academy of Sciences}\ }\textbf
  {\bibinfo {volume} {121}},\ \bibinfo {pages} {e2408717121} (\bibinfo {year}
  {2024})}\BibitemShut {NoStop}%
\bibitem [{\citenamefont {Damascelli}\ \emph {et~al.}(2003)\citenamefont
  {Damascelli}, \citenamefont {Hussain},\ and\ \citenamefont
  {Shen}}]{Damascelli:RMP2003}%
  \BibitemOpen
  \bibfield  {author} {\bibinfo {author} {\bibfnamefont {A.}~\bibnamefont
  {Damascelli}}, \bibinfo {author} {\bibfnamefont {Z.}~\bibnamefont
  {Hussain}},\ and\ \bibinfo {author} {\bibfnamefont {Z.-X.}\ \bibnamefont
  {Shen}},\ }\bibfield  {title} {\bibinfo {title} {Angle-resolved photoemission
  studies of the cuprate superconductors},\ }\href
  {https://doi.org/10.1103/RevModPhys.75.473} {\bibfield  {journal} {\bibinfo
  {journal} {Rev. Mod. Phys.}\ }\textbf {\bibinfo {volume} {75}},\ \bibinfo
  {pages} {473} (\bibinfo {year} {2003})}\BibitemShut {NoStop}%
\bibitem [{\citenamefont {St-Cyr}\ and\ \citenamefont
  {Sénéchal}(2025)}]{StCyr:2025}%
  \BibitemOpen
  \bibfield  {author} {\bibinfo {author} {\bibfnamefont {L.-B.}\ \bibnamefont
  {St-Cyr}}\ and\ \bibinfo {author} {\bibfnamefont {D.}~\bibnamefont
  {Sénéchal}},\ }\bibfield  {title} {\bibinfo {title} {{Effect of the Coulomb
  repulsion and oxygen level on charge distribution and superconductivity in
  the Emery model for cuprates superconductors}},\ }\href
  {https://doi.org/10.21468/SciPostPhysCore.8.2.043} {\bibfield  {journal}
  {\bibinfo  {journal} {SciPost Phys. Core}\ }\textbf {\bibinfo {volume} {8}},\
  \bibinfo {pages} {043} (\bibinfo {year} {2025})}\BibitemShut {NoStop}%
\bibitem [{\citenamefont {{Weber}}\ \emph {et~al.}(2010)\citenamefont
  {{Weber}}, \citenamefont {{Haule}},\ and\ \citenamefont
  {{Kotliar}}}]{Cedric:NatPhys2010}%
  \BibitemOpen
  \bibfield  {author} {\bibinfo {author} {\bibfnamefont {C.}~\bibnamefont
  {{Weber}}}, \bibinfo {author} {\bibfnamefont {K.}~\bibnamefont {{Haule}}},\
  and\ \bibinfo {author} {\bibfnamefont {G.}~\bibnamefont {{Kotliar}}},\
  }\bibfield  {title} {\bibinfo {title} {{Strength of correlations in electron-
  and hole-doped cuprates}},\ }\href {https://doi.org/10.1038/nphys1706}
  {\bibfield  {journal} {\bibinfo  {journal} {Nature Physics}\ }\textbf
  {\bibinfo {volume} {6}},\ \bibinfo {pages} {574} (\bibinfo {year}
  {2010})}\BibitemShut {NoStop}%
\bibitem [{\citenamefont {Weber}\ \emph {et~al.}(2012)\citenamefont {Weber},
  \citenamefont {Yee}, \citenamefont {Haule},\ and\ \citenamefont
  {Kotliar}}]{Weber2011}%
  \BibitemOpen
  \bibfield  {author} {\bibinfo {author} {\bibfnamefont {C.}~\bibnamefont
  {Weber}}, \bibinfo {author} {\bibfnamefont {C.}~\bibnamefont {Yee}}, \bibinfo
  {author} {\bibfnamefont {K.}~\bibnamefont {Haule}},\ and\ \bibinfo {author}
  {\bibfnamefont {G.}~\bibnamefont {Kotliar}},\ }\bibfield  {title} {\bibinfo
  {title} {Scaling of the transition temperature of hole-doped cuprate
  superconductors with the charge-transfer energy},\ }\href
  {http://stacks.iop.org/0295-5075/100/i=3/a=37001} {\bibfield  {journal}
  {\bibinfo  {journal} {EPL (Europhysics Letters)}\ }\textbf {\bibinfo {volume}
  {100}},\ \bibinfo {pages} {37001} (\bibinfo {year} {2012})}\BibitemShut
  {NoStop}%
\bibitem [{\citenamefont {Weber}\ \emph {et~al.}(2010)\citenamefont {Weber},
  \citenamefont {Haule},\ and\ \citenamefont {Kotliar}}]{cedricApical}%
  \BibitemOpen
  \bibfield  {author} {\bibinfo {author} {\bibfnamefont {C.}~\bibnamefont
  {Weber}}, \bibinfo {author} {\bibfnamefont {K.}~\bibnamefont {Haule}},\ and\
  \bibinfo {author} {\bibfnamefont {G.}~\bibnamefont {Kotliar}},\ }\bibfield
  {title} {\bibinfo {title} {Apical oxygens and correlation strength in
  electron- and hole-doped copper oxides},\ }\href
  {https://doi.org/10.1103/PhysRevB.82.125107} {\bibfield  {journal} {\bibinfo
  {journal} {Phys. Rev. B}\ }\textbf {\bibinfo {volume} {82}},\ \bibinfo
  {pages} {125107} (\bibinfo {year} {2010})}\BibitemShut {NoStop}%
\bibitem [{\citenamefont {Cui}\ \emph {et~al.}(2022)\citenamefont {Cui},
  \citenamefont {Zhai}, \citenamefont {Zhang},\ and\ \citenamefont
  {Chan}}]{Cui:Science2022}%
  \BibitemOpen
  \bibfield  {author} {\bibinfo {author} {\bibfnamefont {Z.-H.}\ \bibnamefont
  {Cui}}, \bibinfo {author} {\bibfnamefont {H.}~\bibnamefont {Zhai}}, \bibinfo
  {author} {\bibfnamefont {X.}~\bibnamefont {Zhang}},\ and\ \bibinfo {author}
  {\bibfnamefont {G.~K.-L.}\ \bibnamefont {Chan}},\ }\bibfield  {title}
  {\bibinfo {title} {{Systematic electronic structure in the cuprate parent
  state from quantum many-body simulations}},\ }\href
  {https://doi.org/10.1126/science.abm2295} {\bibfield  {journal} {\bibinfo
  {journal} {Science}\ }\textbf {\bibinfo {volume} {377}},\ \bibinfo {pages}
  {1192} (\bibinfo {year} {2022})}\BibitemShut {NoStop}%
\bibitem [{\citenamefont {Bacq-Labreuil}\ \emph {et~al.}(2025)\citenamefont
  {Bacq-Labreuil}, \citenamefont {Lacasse}, \citenamefont {Tremblay},
  \citenamefont {Sénéchal},\ and\ \citenamefont {Haule}}]{BacqLabreuil:2025}%
  \BibitemOpen
  \bibfield  {author} {\bibinfo {author} {\bibfnamefont {B.}~\bibnamefont
  {Bacq-Labreuil}}, \bibinfo {author} {\bibfnamefont {B.}~\bibnamefont
  {Lacasse}}, \bibinfo {author} {\bibfnamefont {A.-M.~S.}\ \bibnamefont
  {Tremblay}}, \bibinfo {author} {\bibfnamefont {D.}~\bibnamefont
  {Sénéchal}},\ and\ \bibinfo {author} {\bibfnamefont {K.}~\bibnamefont
  {Haule}},\ }\bibfield  {title} {\bibinfo {title} {{Toward an ab initio theory
  of high-temperature superconductors: a study of multilayer cuprates}},\
  }\href {https://doi.org/10.1103/PhysRevX.15.021071} {\bibfield  {journal}
  {\bibinfo  {journal} {Phys. Rev. X}\ }\textbf {\bibinfo {volume} {15}},\
  \bibinfo {pages} {021071} (\bibinfo {year} {2025})}\BibitemShut {NoStop}%
\bibitem [{\citenamefont {{Cui}}\ \emph {et~al.}(2025)\citenamefont {{Cui}},
  \citenamefont {{Yang}}, \citenamefont {{T{\"o}lle}}, \citenamefont {{Ye}},
  \citenamefont {{Yuan}}, \citenamefont {{Zhai}}, \citenamefont {{Park}},
  \citenamefont {{Kim}}, \citenamefont {{Zhang}}, \citenamefont {{Lin}},
  \citenamefont {{Berkelbach}},\ and\ \citenamefont
  {{Chan}}}]{Cui:NatComm2025}%
  \BibitemOpen
  \bibfield  {author} {\bibinfo {author} {\bibfnamefont {Z.-H.}\ \bibnamefont
  {{Cui}}}, \bibinfo {author} {\bibfnamefont {J.}~\bibnamefont {{Yang}}},
  \bibinfo {author} {\bibfnamefont {J.}~\bibnamefont {{T{\"o}lle}}}, \bibinfo
  {author} {\bibfnamefont {H.-Z.}\ \bibnamefont {{Ye}}}, \bibinfo {author}
  {\bibfnamefont {S.}~\bibnamefont {{Yuan}}}, \bibinfo {author} {\bibfnamefont
  {H.}~\bibnamefont {{Zhai}}}, \bibinfo {author} {\bibfnamefont
  {G.}~\bibnamefont {{Park}}}, \bibinfo {author} {\bibfnamefont
  {R.}~\bibnamefont {{Kim}}}, \bibinfo {author} {\bibfnamefont
  {X.}~\bibnamefont {{Zhang}}}, \bibinfo {author} {\bibfnamefont
  {L.}~\bibnamefont {{Lin}}}, \bibinfo {author} {\bibfnamefont {T.~C.}\
  \bibnamefont {{Berkelbach}}},\ and\ \bibinfo {author} {\bibfnamefont
  {G.~K.-L.}\ \bibnamefont {{Chan}}},\ }\bibfield  {title} {\bibinfo {title}
  {{Ab initio quantum many-body description of superconducting trends in the
  cuprates}},\ }\href {https://doi.org/10.1038/s41467-025-56883-x} {\bibfield
  {journal} {\bibinfo  {journal} {Nature Communications}\ }\textbf {\bibinfo
  {volume} {16}},\ \bibinfo {eid} {1845} (\bibinfo {year} {2025})}\BibitemShut
  {NoStop}%
\end{thebibliography}

\end{document}